\documentclass[useAMS,usenatbib]{mn2e}
\usepackage{amssymb}
\usepackage{amsmath}
\usepackage{graphicx} 
\usepackage{multirow}
\usepackage{subfigure} 
\usepackage{aas_macros}
\newcommand{\degree}{\ensuremath{^\circ}}

\newcommand{\parenthnewln}{\right.\\ \left.\quad\quad{}}
\newcommand{\parenthdoublenewln}{\right. \right. \\ \left. \left. \quad\quad{}}
\def\NB6{\texttt{NBODY6}}

\title[The effects of fly-bys on planetary systems]
{The effects of fly-bys on planetary systems}
\author[D. Malmberg et al.]
{Daniel Malmberg$^{1},$ 
Melvyn B. Davies$^{1}$\thanks{E-mail: mbd@astro.lu.se}, Douglas C. Heggie$^{2}$ \\
$^{1}$Lund Observatory, Department of Astronomy and Theoretical Physics, Lund University, Box 43, SE--221 00, Lund, Sweden \\
$^{2}$School of Mathematics and Maxwell Institute for Mathematical Sciences, 
University of Edinburgh, James Clerk Maxwell Building, \\ The Kings Buildings,
Edinburgh EH9 3JZ, UK\\
 }
\begin{document}
\date{Accepted for publication in MNRAS}
\pagerange{\pageref{firstpage}--\pageref{lastpage}} \pubyear{2010}
\maketitle
\label{firstpage}
\begin{abstract}
Most of the observed extrasolar planets are found on tight 
and often eccentric orbits. The high eccentricities are not 
easily explained by planet-formation models, which predict 
that planets should be on rather circular orbits. Here we 
explore whether fly-bys involving planetary systems with properties 
similar to those of the gas giants in the solar system, can 
produce planets with properties similar to the observed 
planets. Using numerical simulations, we show that
fly-bys can cause the immediate ejection of planets, and  
sometimes also lead to the capture of one or more 
planets by the intruder. More common, however, is that 
fly-bys only perturb the orbits of planets, sometimes 
leaving the system in an unstable state. Over time-scales
of a few million to several hundred million years after 
the fly-by, this perturbation can trigger planet-planet scatterings, leading to the ejection of one or more
planets. For example, in the case of the four gas giants of the solar system, the fraction
of systems from which at least one planet is ejected more than doubles in 
$10^8$ years after the fly-by. The remaining planets are often left  on
more eccentric orbits, similar to the eccentricities of
the observed extrasolar planets. We combine our results of how fly-bys
affect solar-system-like planetary systems,
with the rate at which encounters in young stellar clusters occur. For example, we
measure the effects of fly-bys on the four gas giants in the solar system. We find, that
for such systems, between 5 and 15 per cent suffer ejections of planets in
$10^8$ years after fly-bys in typical open clusters. Thus, encounters in young stellar clusters
can significantly alter the properties of any
planets orbiting stars in clusters. As a large fraction of stars which populate the solar
neighbourhood form in stellar clusters, encounters can significantly affect the properties
of the observed extrasolar planets.
\end{abstract}

\section{Introduction} \label{sec:intro}
More than 470 planets have been discovered since the first 
discovery of an extrasolar planet around a main-sequence star 
\citep{1995Natur.378..355M}. The properties of most of these
planets are very different from our solar system, with gas-giant planets on often 
much tighter and more eccentric orbits than seen in the solar system
(http://www.exoplanet.eu).
It is estimated that around 10 per cent of solar mass stars host gas-giant
planets on orbits with semi-major axes less than 5 au \citep{2008PASP..120..531C}.
In particular, the large spread in eccentricities poses a problem for planet formation,
which is believed to occur in discs around young stars
\citep[see, for example,][]{1996Icar..124...62P,1997Sci...276.1836B}. This process will
tend to leave planets on almost circular orbits. 

It is today commonly believed that the observed eccentricities
are caused by so-called planet-planet scattering, first discussed
in \citet{1996Sci...274..954R}. 
Several models have been suggested to explain why 
planetary systems undergo such scattering.
\citet{2008Icar..193..475M} suggest that systems containing two giant
planets, which are migrating in a protoplanetary disc, can perturb each other's
orbits, exciting eccentricities. Such systems are unstable and often result
in the removal of one planet due to ejection, collision or accretion onto the
host star. The remaining planet is left on an eccentric orbit after the gas disc
has dispersed.

It is however also possible that planetary systems evolve past the
disc phase without undergoing scattering. If the planets are too
tightly packed, such systems may become unstable
on time-scales of a few to several 10 million years
\citep[e.g.][]{2008ApJ...686..603J,2008ApJ...686..621F,
2008ApJ...686..580C,2009ApJ...696L..98R}. While 
planet-planet scattering in these types of systems
can explain the observed
eccentricities, migration is needed to explain the observed
semi-major axes \citep[see, for example,][]{2010ApJ...714..194M}.

A somewhat different mechanism to explain the onset
of planet-planet scattering in planetary systems is that 
they are triggered by encounters in the birth environment
of many stars; stellar clusters. 
In such clusters, the high number density of stars causes encounters between stars
to be common. 
So-called exchange encounters can leave planet-hosting stars, which
formed single, in inclined binaries where the so-called Kozai
mechanism can operate \citep{1962AJ.....67..591K}. This can increase
planetary eccentricities, particularly the eccentricity of the outermost
planet, potentially triggering 
planet-planet scattering in multi-planet systems
\citep{2007MNRAS.377L...1M,2009MNRAS.394L..26M}.
However, depending on the binary fraction in stellar clusters,
fly-bys may be much more common than exchange encounters. Essentially,
in clusters with a high binary fraction, exchange encounters are likely
to dominate, while in clusters with a low binary fraction, fly-bys 
are likely play the most important role. Fly-bys can, 
if they occur during the protoplanetary disc stage,
change the frequency of planet formation \citep{2009MNRAS.400.2022F} or the orbital
and physical properties of planets formed \citep{2009A&A...505..873F}.

Fly-bys occurring after the stage of giant planet formation can
excite eccentricities in planetary systems \citep{2004AJ....128..869Z}
and trigger planet-planet scattering in relatively packed systems \citep{2009ApJ...697..458S}.

Young stellar clusters also contain
many young massive stars, whose strong radiation can damage
protoplanetary discs \citep[e.g.][]{2000A&A...362..968A,
2006ApJ...641..504A}.
Our understanding of the birth environment of the solar
system was recently reviewed by \citet{2010arXiv1001.5444A}.
For example, the orbital elements of Sedna, along with the 
chemical composition of objects in the solar system,
suggest that the Sun formed in a cluster with roughly 
$10^3$-$10^4$ stars.

In this paper we measure the effects of fly-bys on planetary systems 
resembling the solar-system, i.e. systems with gas giants on initially
well-spaced and circular orbits. The effects of such fly-bys on both individual
planetary systems, as well as on the population of planetary systems as
a whole, can be significant. Below we summarise the most important findings
from our experiments.
%
\begin{enumerate}
\item  Close fly-bys can lead to the immediate ejection of one or more planets from the host system
\citep[see also][]{1984AJ.....89.1559H,2002ApJ...565.1251H,2006ApJ...640.1086F,2009ApJ...697..458S}. Such planets
may either be left unbound (ionisation), or be captured by the intruder star (exchange). We measure what fraction of 
fly-bys that causes ionisation and exchange to occur in Section 3, and in Table \ref{tab:fracPlSys} we present the rates at which
such encounters occur in stellar clusters with properties representative of the young cluster population in the solar 
neighbourhood. For example, in a cluster with initially 700 stars and half-mass radius of 0.38 pc, 3 per cent of
stars hosting a planetary system similar to the four gas giants of the solar system will suffer a fly-by leading to the immediate
removal of one or more planets from the host star. 

\item Wider fly-bys, which do not immediately remove planets from the host star, may still cause the ejection of planets
on longer time scales. As shown in section 4, the fraction of planetary systems from which at least one planet is
ejected more than doubles in the  $10^8$ years following a stellar encounter. These later ejections are caused
by planet-planet scattering, triggered because the fly-by increased the eccentricities of the planets
\citep[see also][]{2004AJ....128..869Z}, and/or changed their semi-major
axes, leaving the system in an unstable configuration.
 
\item If the intruder in a fly-by is a low-mass star, or even a brown dwarf, an exchange reaction may occur, similar
to what is seen for stellar binaries in clusters \citep[e.g.][]{1975MNRAS.173..729H}. The intruder may then become bound to the
host star, significantly increasing the effect of such fly-bys on planetary systems. We have done 
a large set of scattering simulations of encounters between planetary systems and low-mass intruders, and
find that such encounters can play an important role in the evolution of planetary systems in stellar clusters. 
This shows that a more detailed study of such encounters is warranted and we
will present that in a forthcoming paper.

\item  In general, fly-bys excite the eccentricities of planets, both through direct
scattering off the intruder, and through any subsequent planet-planet scattering. 
 We find that the eccentricity distribution
of planets which have been scattered onto orbits tighter than 4.5 au in fly-bys involving
 planetary systems that initially resembles the solar system, is
similar to that of the observed extrasolar planets (see Fig. 7 and discussion in section 9.2). 
It is important to note that varying the 
initial properties of the planetary systems will significantly affect the properties of the systems post scattering
\citep[see also][]{2008ApJ...686..603J,2009ApJ...696L..98R}. We explore this further
in section 6 and 7.

\item  Planet-planet scattering
 decreases the semi-major axes of one or more planets, while leaving others on either much wider orbits or 
ejected. However, planet-planet scattering in systems resembling the solar system, with
gas giants initially on orbits of 5 au or wider, can only decrease the semi-major axis 
of the inner planet by roughly a factor of two. Hence, the
semi-major axes distribution of such post-scattering systems does not match the observed distribution. 
Planet-planet scattering models thus need a contribution from planetary systems with planets on initially tighter
orbits than those of the gas giants in the solar system in order to reproduce both the observed semi-major axes and eccentricity distributions.
Hence, disc migration, occurring before the stellar encounter, is needed to explain the observed
semi-major axes \citep[see, for example,][]{2010ApJ...714..194M}.

\item  During planet-planet scattering, one or more planets are successively scattered onto wider and wider orbits, until they
become unbound. The remaining planets are scattered onto tighter orbits \citep{Scharf09,Veras09}. During this phase 
it may be possible to observe planets on very wide orbits using imaging techniques. 
It is important to note that this behaviour is independent of what mechanism triggered the planet-planet scattering. As such, 
fly-by induced planet-planet scattering will produce planets on wide orbits. Furthermore, planets which are captured by
the intruder stars in fly-bys are often left on rather wide orbits, and thus could be observed in imaging surveys. We explore
both these mechanisms in Section 8.

\end{enumerate}

In Section 2 we review the encounter rates in young open clusters,
measured using $N$-body simulations \citep{2007MNRAS.378.1207M}.
In particular, we discuss the rate of fly-bys involving solar-mass stars,
as these are most applicable to solar-system-like planetary systems.

In Section 3, we describe the properties of planetary systems immediately
after a fly-by, as measured through both numerical simulations and analytical
calculations. We measure the fraction of systems from which planets are immediately
ejected, and the rate at which planets are captured by the intruder star. We
also measure how important relatively distant fly-bys are in terms of 
moderately exciting the inclinations and eccentricities of planetary systems.

We describe the evolution of the four gas-giants of the solar system
after a fly-by in Section 4. For example, we measure the fraction
of such systems in which planet-planet scattering is triggered within
$10^8$ years after the fly-by. We also discuss the
possibility of predicting the future evolution of such systems immediately
after the fly-by.

In Section 5 we measure the rate at which low-mass intruders are captured 
in encounters with planetary systems, and discuss the implications such encounters
may have on the future evolution of the system.

We proceed to measure the post-fly-by evolution of 
planetary systems slightly different from the 
solar system. In section 7 we go through the 
effects of fly-bys on systems similar to the four gas giants, but where
we have changed the masses of the planets. In section
8 we simulate the evolution after a fly-by, of a planetary system
consisting of 5 gas-giants, generated in simulations of planet 
formation by \citet{1998AJ....116.1998L}.

In section 9 we combine the encounter rates in stellar clusters,
with the measured effects of fly-bys, to understand the effect
on the population of planets in clusters of different sizes.
We also describe some of the properties of this population.

In section 10, we discuss the implications of 
encounters where the intruding star is in fact an
intruding binary. We do not consider the effect of such encounters
here. In our simulations of stellar clusters 
\citep{2007MNRAS.378.1207M} we had a relatively low
primordial binary fraction, and as such these
encounters are not frequent. If we were to increase the primordial
binary fraction, however, we would see many more such encounters,
and we discuss how this would affect our results.
We also, for example, review some of the effects of fly-bys on protoplanetary 
discs, and discuss how changing the primordial binary fraction 
in our clusters can affect our results. We 
summarise our results in section 11.
%
\section{Close encounter rates in young stellar clusters} \label{sec:closeEncRates}
In this section we review our simulations of stellar clusters, and
describe the encounter rates of solar-mass stars.

\subsection{Cluster properties}
We have performed a large set of N-body simulations of young stellar clusters,
with properties representative of the open cluster population in the solar 
neighbourhood \citep[see][for more details]{2007MNRAS.378.1207M}. 
We used the publicly available package NBODY6, which is
a full force-summation direct $N$-body code. A complete description of the
physics included and algorithms used in NBODY6 can be found in \citet{2003gnbs.book.....A}.
During the simulations, we measured the encounter rates of stars and binaries, focussing
especially on fly-bys and so-called exchange encounters. In our simulations we define a fly-by as when
two stars pass each other with in a minimum separation of $r_{\rm min}<1000 \, \mbox{au}. $
In an exchange encounter, a single star
encounters a binary system and replaces one of the stars in the binary. 

In our simulations we consider the evolution of the stellar
clusters after the removal of the primordial gas initially present in the cluster. 
The properties of the clusters were chosen so as to represent the observed
open clusters in the solar neighbourhood. 
At least 10 per cent of stars form in such clusters,
 while the remaining form either in clusters which disperse due to
 gas removal or in small groups or associations
 \citep[e.g.][]{1981PASP...93..712V,
1985ApJ...294..523E,1991MNRAS.249...76B,2001ApJ...553..744A}. 
 Clusters which survive the 
loss of primordial gas nevertheless disperse; 
our simulated clusters have lifetimes of between 400 and 1100
million years.

In this paper, we consider clusters with an
initial half-mass radius, $r_{\rm h}=0.38$ pc and an initial number of stars of $N=150,300,
500, 700$ and $1000$ respectively. In all these simulations, one third of stars were
in primordial binaries. The stellar masses
were drawn from the initial mass function of \citet{1993MNRAS.262..545K} with
a lower mass of $0.2 M_{\odot}$ and upper mass of $5M_{\odot}$. The
initial positions of the stars were chosen such that their distribution followed 
the spherically symmetric \citet{1911MNRAS..71..460P}
model:

\begin{equation}
\rho(\mathbf{r}) = \frac{3 M_{\rm cl}}{4\pi r^3_0} \frac{1}{[1+(r/r_0)^2]^{5/2}},
\end{equation}
for total cluster mass $M_{\rm cl}$.  Here, $r_0$ is a constant, which is 
connected to the half-mass radius of the cluster as
\citep{2003gmbp.book.....H}: $r_{\rm h} \approx 1.305 r_0$.

For each particular cluster type, we performed 10 realisations, performing
a new draw from the IMF and generating a new set of initial positions and
velocities for each realisation.

\subsection{Encounter rates}
 In Fig. \ref{fig:vennAllSingleStars} we show the fractions of initially single stars 
that underwent fly-bys and/or exchange
 encounters in the cluster with $N=700$.
 The results are averaged over 10 realisations. The cluster 
quickly expanded from its initial half-mass radius of $r_{\rm h}=0.38$ 
pc to $r_{\rm h} \sim 2$ pc, whereafter $r_{\rm min}$ was roughly constant
for the remaining lifetime of the cluster. From now on, we will 
refer to this cluster as our reference cluster.
The circle labelled B contains the stars which spent some time
in a binary during the cluster's lifetime, the circle labelled F contains those stars which
underwent a fly-by with other stars.  Finally, the circle labelled S contains those stars which were
single at the end of the cluster's lifetime. About 6.5 per cent of the initially single stars
spent time inside a binary system, while about 73.5 per cent underwent a at least one
fly-by with another star. About 20 per cent of the initially single stars did not spend time
inside a binary or undergo a fly-by. We call such stars {\sl singletons}. A singleton is
defined by us as:

\begin{enumerate}
\item a star which has not formed in a binary,
\item a star which has not later spent time within a binary
system,
\item a star which has not suffered close encounters with other
stars.
\end{enumerate}

A planetary system orbiting a star that is {\sl not} a singleton may have
had its evolution changed due to interactions with other stars.

In Fig. \ref{fig:vennSolarMassSingleStars} we show the interaction rates in our reference
cluster (same cluster as in Fig. \ref{fig:vennAllSingleStars}), but now only for 
stars with $0.8 \, M_{\odot}<m<1.2 \, M_{\odot}$. 
As can be seen in the figure, the fraction of initially single stars which
remain singletons throughout the lifetime of the cluster is lower for solar-mass stars;
only about 15 per cent. 
The fraction of solar-mass stars which spend some time
within a binary system is roughly similar, while the 
fraction which have undergone at least one fly-by is higher (78 per cent). This increase
in the frequency of interactions is due in part to the effects of mass segregation, which draws the heavier
solar-like stars into the denser central regions of the cluster.

In Fig. \ref{fig:nFBperStar} we plot the distribution of the number of fly-bys 
experienced, $n_{\rm fb}$, for the initially-single solar-mass stars that
have undergone fly-bys in our reference cluster. About 75 per cent
of them have experienced two or more fly-bys. The mean 
number of fly-bys per star is in fact four.  Stars which have undergone several fly-bys will, on average, 
have closer encounters with other stars than systems only undergoing a single flyby.
The cross section for two stars, having a relative velocity at infinity
of $v_\infty$, to pass within a minimum distance $r_{\rm min}$ is given by

\begin{equation} \label{eq:CrossSection}
\sigma = \pi r_{\rm min}^2 \left( 1 + {v^2 \over v_\infty^2} \right)
\end{equation}

\noindent where $v$ is the relative velocity of the two
stars at closest approach in a parabolic encounter {\it i.e.\ }
$v^2 = 2 G (m_1 + m_2)/r_{\rm min}$, where $m_1$ and $m_2$ are the masses
of the two stars.
The second term is due to the attractive gravitational force, and
is referred to as gravitational focussing.
In the regime where $v \ll v_\infty$ (as might be the case in
galactic nuclei with extremely high velocity dispersions), we recover
the result, $\sigma \propto r_{\rm min}^2$. However, if $v \gg v_\infty$
as will be the case in systems with low velocity dispersions, such as
young clusters, $\sigma \propto r_{\rm min}$.
Thus a star that has had a number of fly-bys, $n_{\rm fb}$, within 1000 au  will on average have passed
within a mean minimum distance, ${\langle r_{\rm min} \rangle}_{\rm min}$, of:

\begin{equation}
{ \langle r_{\rm min} \rangle}_{\rm min} = 1000 \, \mbox{au}/n_{\rm fb}. 
\end{equation}

As the mean number of fly-bys in our reference cluster is four, we thus
find  $ {\langle r_{\rm min} \rangle}_{\rm min} = 250$ au.

In Fig. \ref{fig:fracFBclusterAge} we plot the distribution of when fly-bys occur, $t_{\rm fb}$
($r_{\rm min} < 1000$ au) as a function of cluster age. As can be seen in the figure, 
the evolution is very similar
for all clusters. On average, half of the fly-bys occur within the first 10 million years
This time-scale is in principle comparable to the gas-giant planet formation time-scale,
but as our simulations only treat the post-gas-loss life of clusters, 
gas-giant planets have already formed at the start of our simulations.

{\sl 
\begin{figure}
\begin{center}
\resizebox{8truecm}{!}{\includegraphics{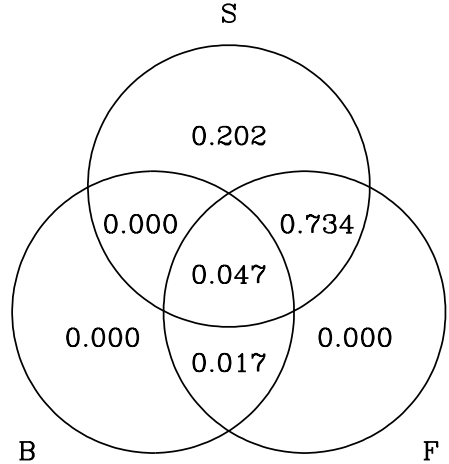}}
\caption{Venn diagram of the stars which were initially single in our reference
cluster ($N = 700$ and $r_{\rm h,initial} = 0.38$ pc). 
The numbers are the average result from 10 realisations.
The upper circle contains stars which were single at the end of the run (S), 
the lower left-hand circle
contains the stars which were in a bound system (i.e. triple or binary) during the 
lifetime of the cluster
(B) and the lower right-hand circle contains the stars which had experienced
a fly-by (F) (defined as when two stars pass within 1000 au of
each other).}
\label{fig:vennAllSingleStars}
\end{center}
\end{figure}

%

\begin{figure}
\begin{center}
\resizebox{8truecm}{!}{\includegraphics{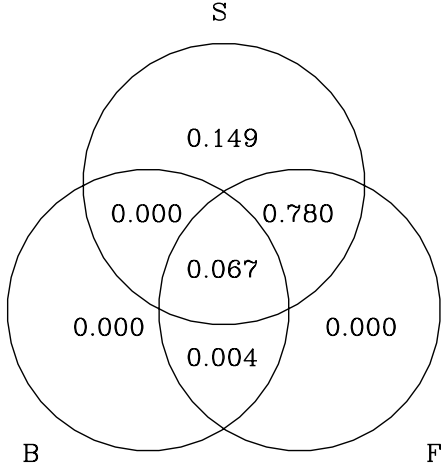}}
\caption{Venn diagram of the stars which were initially single in our reference
cluster ($N = 700$ and $r_{\rm h,initial} = 0.38$ pc) and had mass between
$0.8 M_{\odot}$ and $1.2 M_{\odot}$ .
The numbers are the average result from 10 realisations.
The upper circle contains stars which were single at the end of the run (S), 
the lower left-hand circle
contains the stars which were in a bound system (i.e. triple or binary) during the 
lifetime of the cluster
(B) and the lower right-hand circle contains the stars which had experienced
a fly-by (F) (defined as when two stars pass within 1000 au of
each other). }
\label{fig:vennSolarMassSingleStars}
\end{center}
\end{figure}
\begin{figure}
\begin{center}
\resizebox{8truecm}{!}{\includegraphics{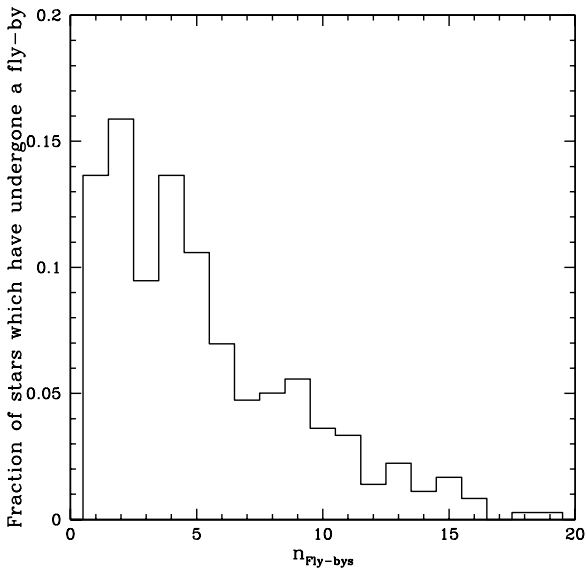}}
\caption{The distribution of number of fly-bys, $n_{\rm fly-bys}$, per star
for the initially single stars with mass $0.8<m<1.2 M_{\odot}$ that experienced
fly-bys in our reference cluster ($N = 700$ and $r_{\rm h,initial} = 0.38$ pc).
The average number of fly-bys per star in the cluster is 4.}
\label{fig:nFBperStar}
\end{center}
\end{figure}
%
\begin{figure}
\begin{center}
\resizebox{8truecm}{!}{\includegraphics{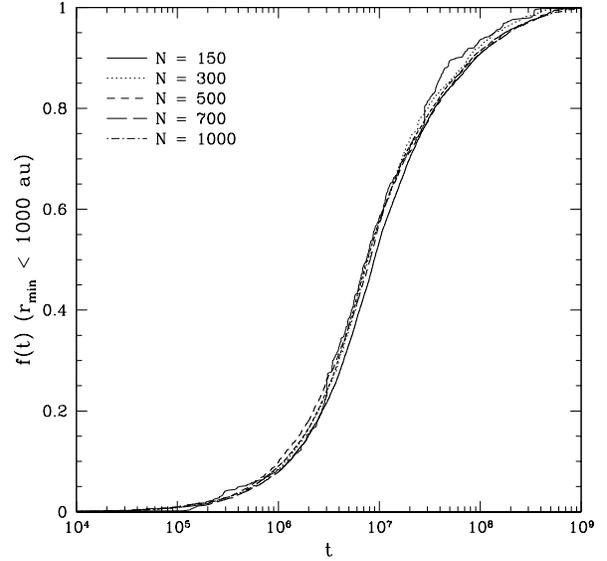}}
\caption{The fraction of all encounters which have occurred as function
of cluster age, averaged over the 10 realisations for each cluster. 
Only clusters with initial half-mass radii of 0.38 pc are included. 
These clusters have radii of about 2-4 pc during most of their lifetime.}
\label{fig:fracFBclusterAge}
\end{center}
\end{figure}
 }

%
%
%
\section{The properties of planetary systems immediately after a fly-by} \label{sec:instantEffectOfCloseEnc}
We begin here by describing in detail the effects of fly-bys on a planetary system consisting 
of the four gas-giant planets of the solar system (4G).
The immediate effect of a fly-by depends crucially on the minimum separation
between the intruder star and the planetary-system-host star during the encounter, $r_{\rm min}$.
 We divide fly-bys into two different regimes, depending on $r_{\rm min}$; the strong regime
($r_{\rm min}<100$ au) and the the weak regime ($100<r_{\rm min}<1000$ au).

To analyse the encounters we use a combination of direct numerical simulations 
(in the strong regime) and  analytic theory (in the weak regime).
In all our simulations, we set the relative velocity of the two stars, $v_{\infty} = 1$ km/s,
which is the typical velocity dispersion in young stellar clusters.

To perform the numerical simulations, we use the
Bulirsch--Stoer algorithm, as implemented in the MERCURY6 package 
\citep{1999MNRAS.304..793C}.

\subsection{The strong regime: $r_{\rm min} < 100$ au}
\begin{figure}
\begin{center}
\includegraphics[scale=0.4]{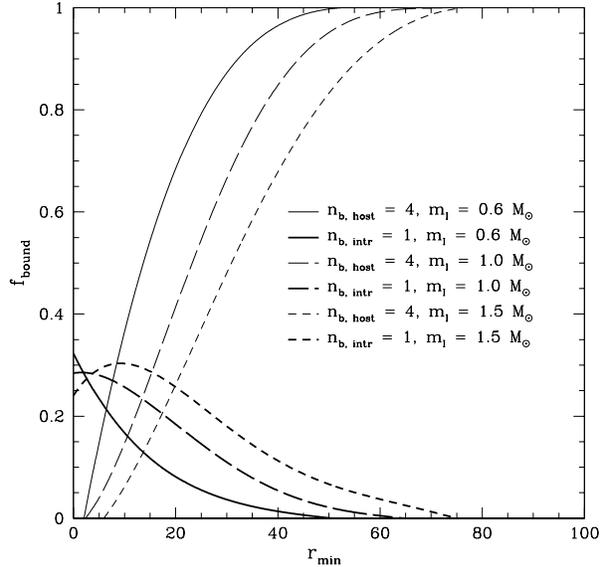}
\caption{The probability that no planets are ejected or captured (thin lines) 
and that one planet is captured by the intruder star (thick lines) immediately
after the fly-by as a function of $r_{\rm min}$ for encounters involving the four 
gas giants. The solid lines are for encounters where $m_{\rm I}=0.6 M_{\odot}$, 
the long-dashed lines are for encounters where $m_{\rm I}=1.0 M_{\odot}$ and 
the short-dashed lines are for encounters where $m_{\rm I}=1.5 M_{\odot}$.}
\label{fig:nBoundrMin}
\end{center}
\end{figure}

Close fly-bys 
cause significant changes in the orbits of planets. Often, they 
lead to the immediate ejection of planets or an exchange, 
in which planets become bound
to the intruder star. We have performed
$6 \times 10^5$ scattering experiments of fly-bys involving the four
gas-giant planets of the solar system, with  $r_{\rm min}<100$ au 
and intruder-star mass, $m_{\rm I} = 0.6,1.0$ and $1.5 M_{\odot}$.

In Fig. \ref{fig:nBoundrMin} we plot the probability that four 
planets remain bound to the host star immediately after the 
fly-by (thin lines) and the probability that one planet
is captured by the intruder star (thick lines), as a function 
of $r_{\rm min}$.
The semi-major axis of the outermost planet, Neptune, is 30 au. 
The probability that all four planets remain bound to the intruder star
(thin lines) reaches unity at between two and three times the semi-major axis of the outermost 
planet, depending on the mass of the intruder star. For example, in
fly-bys with $r_{\rm min} <100$ au and $m_{\rm I}=0.6 M_{\odot}$, 15 per cent of fly-bys 
lead to one or more planets becoming unbound from the host star.
Hence, in  fly-bys where
$r_{\rm min}$ is comparable in size to the semi-major axes of the planets' orbits, 
the result is often to immediately unbind one or more planets.
Sometimes planets are ejected from the system and left as free-floating planets,
but it is not uncommon that planets instead are captured by the intruder star, and
hence orbit it after the fly-by. In the latter case, the most likely outcome is that only
one planet is captured (thick lines in Fig. \ref{fig:nBoundrMin}).

\subsection{The weak regime: $100 < r_{\rm min} < 1000$ au}
\begin{figure}
\begin{center}
\includegraphics[scale=0.4]{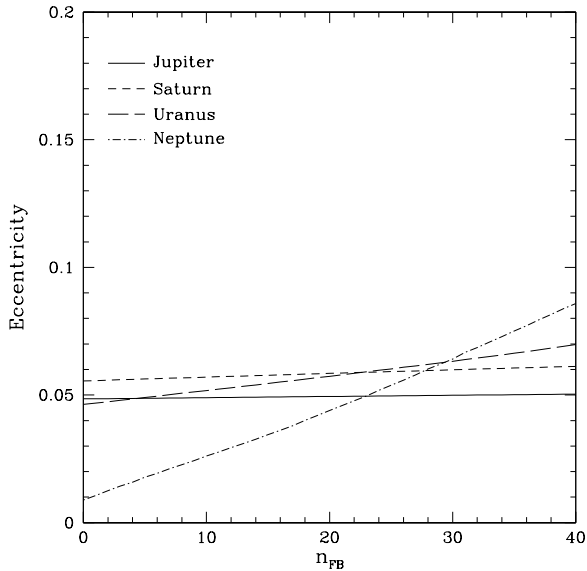}
\caption{The eccentricities of the the four gas giants of the solar system as function of number of
fly-bys, $n_{\rm fb}$. The minimum separation between the host star and the intruder star was
chosen randomly between 100 and 1000 au.}
\label{fig:nFB_ecc4G_weakRegime}
\end{center}
\end{figure}

In encounters with $r_{\rm min}>100$ au we can measure the 
change in eccentricities and inclinations of the planets
analytically \citep{1996MNRAS.282.1064H}.
Three different cases must be considered, depending on the initial 
eccentricity of the planet (circular or eccentric) and on $r_{\rm min}$.

We have calculated the the average change in the eccentricity of the gas giant planets in the
solar system with number of fly-bys, $n_{\rm fb}$ 
using equation (\ref{eq:eccOrbit}), (\ref{eq:circOrbitDist}) and (\ref{eq:circOrbTight}) 
from Appendix A.
We assume that the intruder is randomly oriented with respect to the planets and 
has $m_{\rm I}=1.0 M_{\odot}$. We plot the result in Fig. \ref{fig:nFB_ecc4G_weakRegime}. 
A large number of fly-bys is required to significantly change the mean eccentricities of the 
planets. Also, as the cross-section for encounters scales with $r_{\rm min}$, a star will experience
at least one encounter with $r_{\rm min}<100$ au for every 10 encounters it experiences 
with $100<r_{\rm min}<1000$ au. It is thus unlikely that fly-bys in the weak
regime trigger planet-planet scatterings in planetary system similar to the solar system. 
They can, however, be important as a ``heating mechanism" of planetary systems. For example,
if we assume that the solar system initially started out circular and co-planar, we can ask 
the question ``How many fly-bys, $n_{\rm fb}$ would be required to increase the values of the
planets' eccentricities and inclinations to their current values?" 
A complication here is that planet-planet
interactions work to transfer angular momentum between planets,
continuously changing the eccentricities and inclinations of the planets.
Hence, any change in the inclination and eccentricities of the planets will propagate through
the system \citep[see, for example,][]{2004AJ....128..869Z}.

To estimate the number of fly-bys, $n_{\rm fb}$, required to take an initially
co-planar and circular version of the solar system and give it similar eccentricities
and inclinations to the present day solar system, we have chosen to calculate the 
so-called angular momentum deficit of the system, as a function of $n_{\rm fb}$.

The angular momentum deficit (AMD) in a system is the additional angular momentum
present in it due to the non-circularity and non-co-planarity of the planetary orbits. It can
be thought of as the amount of non-linearity present in the system.
It is defined as \citep{2000PhRvL..84.3240L}:

\begin{equation} \label{eq:AMD}
AMD = \sum^{n_{\rm p}}_{k=1} \Lambda_{\rm k} (1-\sqrt{1-e_{\rm k}^2} \cos{i_{\rm k}})
\end{equation}
where $e_{\rm k}$ and  $i_{\rm k}$ are the eccentricity and inclination of the $k^{\rm th}$ planet,
and $\Lambda = m_{\rm k} \sqrt{\mu a_{\rm k}}$, where $m_{\rm k}$ and $a_{\rm k}$ are
the mass and semi-major axis of the $k$th planet. 
Starting out with a co-planar system with planets on initially circular orbits
we find that  15  fly-bys with solar-mass intruders with $100<r_{\rm min}<1000$,
is enough to give the system its present AMD value.

\section{Post-encounter evolution of planetary systems} \label{sec:postEncounterEvolution}
\begin{figure}
\begin{center}
\subfigure[]{\includegraphics[scale=0.4]{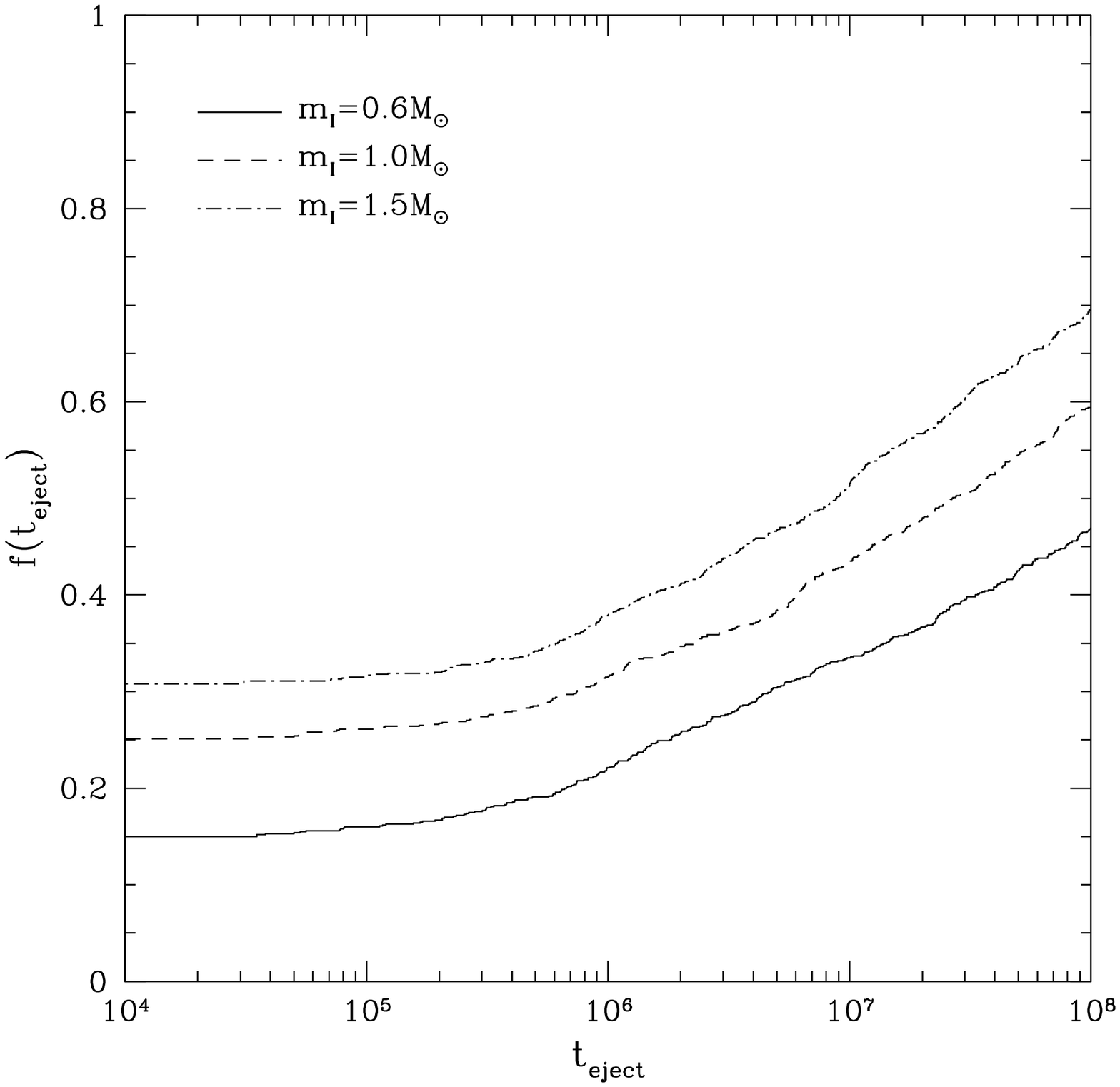}}
\subfigure[]{\includegraphics[scale=0.4]{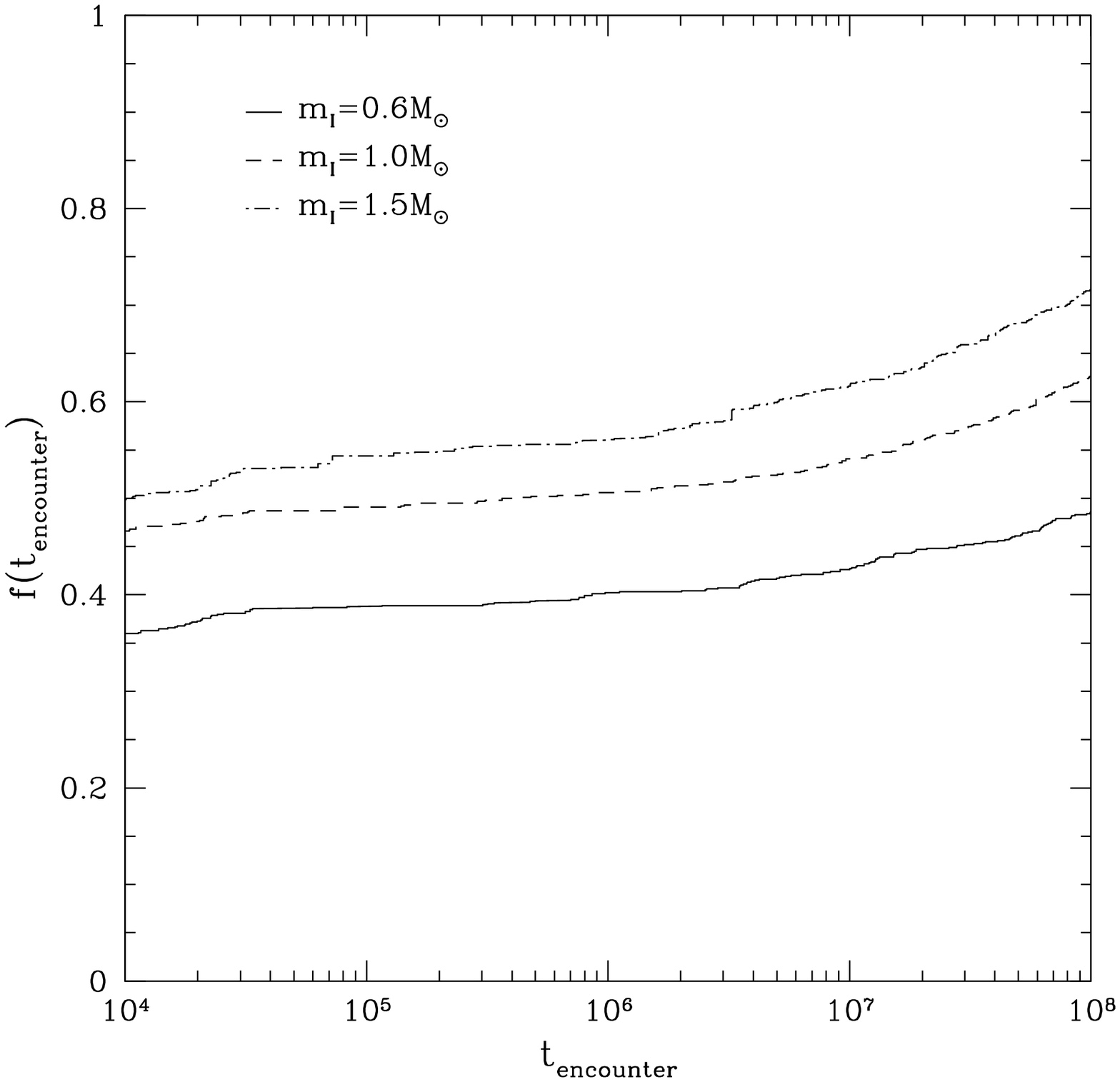}}
\caption{The evolution of the four gas giants of the solar system, after a fly-by
with $r_{\rm min}<100$ au. The periastron passage of the intruder star occurred
at $t=0$. In panel a) we plot the fraction of systems, $f(t_{\rm eject}$), from which at
least one planet has been ejected within, $t_{\rm eject}$ years after the fly-by and 
in panel b) we plot the fraction of systems, $f(t_{\rm encounter}$), in which at
least two planets have passed within one mutual Hill radius of each other as
a function of time, $t_{\rm encounter}$ after the fly-by.}
\label{fig:tInt4G}
\end{center}
\end{figure}

\begin{table*}
\caption{Overview of the evolution of the four gas giants of the solar system after 
a fly-by, divided into different categories
depending on the state of the system immediately after the fly-by. In column 2 we list the
categories and in column 3 we list the fraction of all systems which belongs to a
given category. In column 4, 5, 6 and 7 we divide the systems from each category
into four bins, depending on at what time after the fly-by the first planet was ejected from the
system, $t_{\rm eject}$. In column 4 are systems with $t_{\rm eject} < 10$ Myr, in column
5 systems with  $t_{\rm eject} < 30$ Myr, in column 6 systems with  $t_{\rm eject} < 100$ Myr
and in column system are systems from which no planet was ejected in $10^8$ years.
Its worth pointing out, that for systems in which a planet is ejected immediately in the fly-by,
column 4 and 5 are by definition 0, as all such planets have $t_{\rm eject} = 0$. Furthermore, the
fraction of systems which falls into the categories ``Kozai immediately after fly-by", ``AMD unstable 
immediately after fly-by" and
 ´´Kozai before close encounter" are in general rather small the uncertainty in these values is
 rather large, especially in column 3-6. For example, in the simulations with $m_{\rm I}=0.6M_{\odot}$,
  only 1.8 per cent of systems, or 18 simulations, fall into the category
´´AMD unstable immediately after fly-by", with 3 systems that have $t_{\rm eject}<10^7$ years (column 3),
2 systems with $10^7<t_{\rm eject}<3\times10^7$ (column 4), 6 systems with
 $3\times10^7<t_{\rm eject}<10^8$ (column 4) and finally 7 systems which are
 AMD unstable immediately after the fly-by, but from which no planets are ejected
 within 100 Myr after the fly-by (column 6). }
\begin{center}
\begin{tabular}{|c|l|c|c|c|c|c|}
 \hline
$m_{\rm I}$ &  & Situation immediately  &\multicolumn{3}{l}{$t_{\rm eject}$ (Myr)} &  No planet ejection\\
& & after flyby  & 10 & 30 & 100 &  after 100 Myr \\
 \hline
$0.6M_{\odot}$ & Immediate ejection:		&  0.150 & 1.00 & 0.00 & 0.00 & 0.00 \\
& Orbit crossing immediately after fly-by:          	&  0.262 & 0.62 & 0.18 & 0.20 & 0.00 \\
& Kozai immediately after fly-by:                   	&  0.010 & 0.40 & 0.00 & 0.10 & 0.50 \\
& AMD unstable immediately after fly-by:         	&  0.018 & 0.17 & 0.11 & 0.33 & 0.39 \\
& Kozai before close encounter:    			&  0.004 & 0.25 & 0.00 & 0.00 & 0.75 \\
& Other (unstable):   						&  0.037 & 0.10 & 0.22 & 0.68 & 0.00 \\
& Other(stable):      						&  0.519 & 0.00 & 0.00 & 0.00 & 1.00 \\
 \\
$1.0M_{\odot}$ & Immediate ejection: 		&  0.251 & 1.00 & 0.00 & 0.00 & 0.00 \\
& Orbit crossing immediately after fly-by:     	&  0.289 & 0.58 & 0.20 & 0.22 & 0.00 \\
& Kozai  immediately after fly-by:  			&  0.012 & 0.25 & 0.08 & 0.17 & 0.50 \\
& AMD unstable immediately after fly-by:       	&  0.015 & 0.00 & 0.27 & 0.27 & 0.46 \\
& Kozai before close encounter:    			&  0.001 & 1.00 & 0.00 & 0.00 & 0.00 \\
& Other (unstable):   						&  0.039 & 0.00 & 0.15 & 0.85 & 0.00 \\
& Other(stable):      						&  0.393 & 0.00 & 0.00 & 0.00 & 1.00 \\
\\
$1.5M_{\odot}$ & Immediate ejection: 		&  0.308 & 1.00 & 0.00 & 0.00 & 0.00 \\
& Orbit crossing immediately after fly-by:   	&  0.304 & 0.58 & 0.20 & 0.22 & 0.00 \\
& Kozai immediately after fly-by:             		&  0.022 & 0.23 & 0.14 & 0.08 & 0.55 \\
& AMD unstable immediately after fly-by:  	&  0.031 & 0.06 & 0.19 & 0.35 & 0.40 \\
& Kozai before close encounter:    			&  0.005 & 0.40 & 0.00 & 0.00 & 0.60 \\
& Other (unstable):   						&  0.050 & 0.16 & 0.30 & 0.54 & 0.00 \\
& Other(stable):      						&  0.280 & 0.00 & 0.00 & 0.00 & 1.00 \\
 \hline
\end{tabular}
\end{center}
\label{tab:stabilityCriteria}
\end{table*}%

In this section we consider the  evolution of the planetary systems
on the longer time-scale after fly-bys. 
In multi-planet systems, planet-planet interactions 
will cause evolution of the planets' orbital elements.
The effect of most fly-bys (except the very tight ones) is to only change the 
angular momenta of the planets, with the largest change for the outermost
planet. While this change may not immediately trigger planet-planet scattering,
redistribution of angular momentum between the planets (e.g.
planet-planet interactions) may
cause the eccentricities of one or more planets to grow over time, 
triggering planet-planet scattering.

\subsection{The long-term evolution of the four gas giants of the solar system after a fly-by}
We begin here by describing the long-term evolution of planetary systems
consisting of the four gas giants of the solar system that have undergone
fly-bys with $r_{\rm min} < 100$ au. 

We have performed three sets of simulations, with $m_{\rm I} = 0.6, 1.0$ and 
$1.5 M_{\odot}$ respectively).
For each set we perform 1000 realisations, between which we randomly vary the
orientation of the intruder star's orbit. 
We vary $r_{\rm min}$ linearly between 0 and 100 au, which is consistent with 
the encounter rates we measure
in our simulations of young stellar clusters (Section \ref{sec:closeEncRates}).
In each simulation we measure the time until the first close encounter
between two planets occur, $t_{\rm encounter}$, and the time until one planet is ejected
$t_{\rm eject}$. A close encounter is defined as
when two planets pass within one mutual Hill radii, $R_{\rm H}$, of each 
other, where:

\begin{equation} \label{eq:mutualHill}
R_{\rm H}= \frac{a_1 + a_2}{2} \left(  \frac{m_1 + m_2}{3M_{\rm H}}\right)^{1/3} 
\end{equation}
where $m_1$ and $m_2$ are the the masses of the planets, $a_1$ and
$a_2$ are their semi-major axes and  $M_{\rm H}$ is the mass of the host star.

In Fig. \ref{fig:tInt4G}a we plot the fraction of systems from which at least
one planet has been ejected, $f(t_{\rm eject})$, as a function of time after the fly-by. 
The three different lines correspond
to different intruder star masses (solid line: $m_{\rm I} = 0.6M_{\odot}$, 
dashed line: $m_{\rm I} = 1.0 M_{\odot}$ and
dash-dotted line: $m_{\rm I} = 1.5 M_{\odot}$). The fraction of systems in
which a planet is ejected immediately after the fly-by increase with $m_{\rm I}$. For
$m_{\rm I} = 0.6, 1.0$ and $1.5 M_{\odot}$ 15, 25 and 31 per cent of systems suffer an immediate
ejection.
The fraction of systems from which at least
one planet has been ejected then grows with time, reaching values of
0.47 ($m_{\rm I} = 0.6M_{\odot}$), 0.59 ($m_{\rm I} = 1.0M_{\odot}$) 
and 0.69 ($m_{\rm I} = 1.5M_{\odot}$) at $10^8$ years after the fly-by. 
Hence, planet-planet scattering, triggered due to the fly-by, more than
doubles the fraction of systems from which at least one planet is ejected
in $10^8$ years for the 4G system. 

In Fig. \ref{fig:tInt4G}b we plot the fraction of systems in which two planets
have undergone a close encounter with each other, $f(t_{\rm encounter})$.
We define a close encounter as when two planets pass within one mutual
Hill radius of each other (see equation \ref{eq:mutualHill}).
We find that $f(t_{\rm encounter})$ is always significantly larger than 
$f(t_{\rm eject})$, implying that the ejection of a planet,
 unless it occurs immediately after the fly-by, is preceded by a phase
 of planet-planet scatterings.
As $f(t_{\rm encounter})$ is larger than $f(t_{\rm eject})$ at $t=10^8$ years
we can also predict that the fraction of systems from
which at least one planet has been ejected will continue to increase for $t>10^8$
years.

\subsection{Predicting the long-term effects of fly-bys} \label{sec:PredictingFBs}
While one in principle could simulate the long-term evolution of planetary systems
which have undergone fly-bys for a large range of planetary system architectures
and fly-by properties, it would be greatly beneficial if one could predict the
future evolution immediately after the fly-by.

We have broken down the post-encounter systems from our simulations
of the four gas giants into seven different categories,
depending both on their properties immediately after the fly-by and on
their long-time evolution.
The systems are put into only one category; the first one in the
following lists whose criteria they fulfil.

\begin{enumerate}

\item {\sl Immediate ejection:} planetary systems in which at least one planet became unbound
from the host star immediately after the fly-by,

\item {\sl Orbit crossing immediately after fly-by:} planetary systems in which the orbits of at least two
planets cross immediately after the fly-by,

\item {\sl Kozai immediately after fly-by:} planetary systems in which the mutual inclination of at least
two planets is between $39.2\degree$ and $140.8\degree$ immediately after the fly-by, 

\item {\sl AMD unstable immediately after fly-by:} planetary systems in which the angular momentum deficit (AMD)
immediately after the encounter is such that the orbits of two planets may cross,

\item {\sl Kozai before close encounter:} planetary systems that become unstable in $10^8$
years in which the mutual inclination between at least two planets attains a value between
 $39.2\degree$ and $140.8\degree$ before the onset of planet-planet scattering,

\item {\sl Other (unstable):} planetary systems which do not fulfil any other criteria,
but from which at least one planet is ejected within $10^8$ years after the fly-by,

\item {\sl Other (stable):} planetary systems which do not fulfil any of the above
criteria and are stable for at least $10^8$ years after the fly-by.

\end{enumerate}

These criteria are described in further detail below.

If the mutual inclination between two planets is between $39.2\degree$ and $140.8\degree$
the so-called Kozai mechanism \citep{1962AJ.....67..591K} operates. The effect of the Kozai
mechanism is to cause the eccentricity and inclination of planets to oscillate,
triggering planet-planet scatterings in multi-planet systems \citep{2007MNRAS.377L...1M,
2009MNRAS.394L..26M}. 

While the intruder star in a fly-by will not by itself cause the Kozai mechanism to operate,
it may leave the system in a state where the mutual inclination between two or more
planets is between $39.2\degree$ and $140.8\degree$. The Kozai
mechanism will then operate, potentially triggering planet-planet scatterings
\citep{2008ApJ...678..498N}. 

It is also possible that planet-planet interactions, which change both the eccentricities
and inclinations of planets with time, increase the mutual inclination between two
planets such that it reaches values between $39.2\degree$ and $140.8\degree$ even
if this is not the case immediately after the fly-by. For systems which do not fulfil the
criteria of category {\sl (i)-(iv)}, but in which planet-planet scatterings occur within $10^8$
years after the fly-by, we check whether this happens, and if so, the systems are put into
category {\sl (v)}.

The angular momentum deficit (AMD, see Equation \ref{eq:AMD}) in a system is the additional angular momentum
present in it due to the non-circularity and non-co-planarity of the planetary orbits. It can
be thought of as the amount of non-linearity present in the system.
We use the AMD to
calculate whether the planetary system may become unstable, by checking if any pair
of planets can achieve high enough eccentricities for their respective orbits
to cross, if all the available AMD is put into these two planets \citep[see][for more details]{2000PhRvL..84.3240L}.

In column 3 of Table \ref{tab:stabilityCriteria} we list the fraction of systems in our simulations
of fly-bys belonging to each of the
categories listed above. In columns 4-6 we list the fraction of systems
from which at least one planet is ejected within $10^7$, $3 \times 10^7$
and $10^8$ years respectively after the fly-by. Finally, in column 7, we list the fraction of systems
from which no planets are ejected in $10^8$ years after the encounter.

We can now in more detail analyse the post-encounter behaviour of the planetary
systems. For example, in encounters with an intruder of mass $0.6M_{\odot}$,
at least one planet is ejected immediately after the fly-by in 15 per cent of
the encounters. An additional 26.2 per cent of all systems are left in a state
where the orbits of at least two planets cross. Thus, in total, 41.2 per cent
of systems fall into categories {\sl (i)} and {\sl (ii)}. A total of 47 per cent
of systems eject at least one planet in $10^8$ years after the fly-by.
Hence, about 88 per cent of the systems which become unstable in $10^8$ 
years after the fly-by, fall into categories {\sl (i)} and {\sl (ii)}.
This behaviour is similar also for fly-bys with intruders of mass 
$1.0 M_{\odot}$ and $1.5M_{\odot}$.

It may seem surprising that some systems left with orbit-crossing planets do not 
eject any planets until 100 Myr after the flyby encounters. In many cases, planets
left on eccentric orbits are also rather inclined to the original plane of the unperturbed
planetary system, so even if orbits cross, close encounters between two planets
may be extremely rare events.
 Only a small fraction of systems
that later become unstable are left in such a state that the Kozai mechanism would 
operate or that they are AMD unstable. About 4-5 per cent of all systems fall into the category
{\sl unstable (other)}.

 Given that the vast majority of systems which eject planets fall
 into categories  {\sl (i)} and {\sl (ii)}, we may use the list of post-flyby categories
 to predict which systems will eject planets. 
It is very useful that we are able to predict the post-flyby  evolution of a system containing 
  the four gas-giants, as it reduces the computational effort needed to understand
the effect of fly-bys. As we shall in section 6 however, 
predicting the post-fly-by evolution of planetary systems which are inherently
less stable than the four gas giants is a more difficult task.

\section{Capture of low-mass intruders}
\begin{table*}
\caption{The calculated cross-sections (column 2-4) that
encounters between a sub-stellar mass object and 
a Jupiter-mass planet orbiting a solar-mass star with
semi-major axis of 30 au, leads to the formation of a transient 3-body system, 
called a resonant interaction, or to a prompt exchange.
For intruder mass $m_{\rm I} = 30 M_{\rm J}$ and $v_{\infty} = 1.0$ km/s,
the encounter sometimes lead to ionisation, where all the
three objects become unbound.
We have calculated the cross-section for three
different values of $m_{\rm I}$: 3, 10
and $30M_{\rm J}$, and two different values of $v_{\infty}$: 
0.5 km/s (top panel) and 1.0 km/s (lower panel).
In column 5 we list the fraction of such encounters, $f_{\rm e}$ which
occur in our reference cluster ($N=700$ stars and $r_{\rm h,i}=1000$ au).
The cross-sections, $\sigma$ has been scaled to give 
the dimensionless quantity ($\sigma_{\rm scaled}=v^2 \sigma/\pi a_{\rm p}^2$), 
where $a_{\rm p}$ is the semi-major axis
of the planet and $v=v_{\infty}/v_{\rm crit}$.}
\begin{center}
\begin{tabular}{|l|c|c|c|c|}
\hline
Type               						& $m_{\rm I} = 3 m_{\rm J}$ 		& $m_{\rm I} = 10 m_{\rm J}$ 			&  $m_{\rm I} = 30 m_{\rm J}$ 			 & $f_{\rm e}$			\\
\hline
\multicolumn{4}{l}{$v_{\infty} = 0.5$ km/s}									 																				\\
Resonant interaction, intruder ejected   	& 8.0$\pm$1.3 $\times 10^{-2}$   	& 2.1$\pm$0.3 $\times 10^{-1}$ 		&   7.3 $\pm$0.5 $\times 10^{-1}$  		& $4.1 \times 10^{-2}$	\\
Resonant interaction, planet ejected  	& 1.0$\pm$0.4 $\times 10^{-2}$  	& 0.7$\pm$0.2 $\times 10^{-1}$  		&   5.6 $\pm$0.5 $\times 10^{-1}$  		& $3.2 \times 10^{-2}$	\\
Prompt exchange, planet ejected   		& 0.0	    						& 0.0     							&   1.5 $\pm$0.3 $\times 10^{-1}$ 		& $1.3 \times 10^{-2}$	\\
Prompt exchange, host star ejected   	& 0.0    						& 0.0    							&   0.0  							& 0					\\
Ionisation, system unbound   			& 0.0    						& 0.0    							&   0.0  							& 0					\\
$v$           							& 0.159 						& 0.291 							&   0.504 												\\
\\
\multicolumn{4}{l}{$v_{\infty} = 1.0$ km/s} 																													\\
Resonant interaction, intruder ejected 	& 9.8$\pm$2.0 $\times 10^{-3}$	& 1.8$\pm$0.6 $\times 10^{-2}$ 		&  0.0							&  $2.0 \times 10^{-3}$	\\
Resonant interaction, planet ejected 	& 3.0$\pm$1.0  $\times 10^{-3}$	& 1.4$\pm$0.5 $\times 10^{-2}$ 		&  0.0 							&  $1.5 \times 10^{-3}$	\\
Prompt exchange, planet ejected  		& 0.0     						& 8.8$\pm$5.0 $\times 10^{-4}$  		& 1.4$\pm$0.2 $\times 10^{-1}$		&  $6.5 \times 10^{-3}$	\\
Prompt exchange, host star ejected  	& 0.0     						& 0.0    							& 1.5$\pm$0.8 $\times 10^{-2}$		&  $2.0 \times 10^{-3}$	\\
Ionisation, system unbound   			& 0.0    						& 0.0   							& 5.2$\pm$5.0 $\times 10^{-3}$		&  $1.2 \times 10^{-3}$	\\
$v$                                   				& 0.318     					& 0.582    							& 1.007 												\\
\hline
\end{tabular}
\end{center}
\label{tab:BDcrossSec}
\end{table*}%

In this section we consider the effects of encounters involving brown dwarfs and 
free--floating planets.  Such encounters are different from fly-bys by
stellar-mass intruders,  as the intruder may be captured. 
These sorts of encounters are rather similar in nature to encounters
between single stars and hard stellar binaries - that is, encounters where the binding energy of
the binary exceeds the kinetic energy of the incoming third star. Ultimately, such encounters
result in the ejection of one of the three stars, leaving the binary more tightly bound. In the case
of encounters with a star possessing a planetary system, the encounter can be somewhat
complicated given the larger number of bodies involved. If the incoming planet is captured,
even temporarily, close encounters between it and other planets in the system can occur,
leading to the ejection of planets, or at least leaving them on altered orbits.

Roughly speaking, a transient 3-body system can only form if the binding energy of the planetary system 
is larger than the kinetic energy of the intruder star. We can write the
critical velocity, $v_{\rm crit}$, for which the total energy of the intruder and planetary
systems equals zero, as:

\begin{equation} \label{eq:vCrit}
v_{\rm crit} = \left( \frac {G m_{\rm H} m_{\rm p}}{\mu a_{\rm p}} \right)^{1/2} 
\end{equation}
where $m_{\rm H}$ is the mass of the planet host star, $m_{\rm p}$ is the mass of the planet,
$m_{\rm I}$ is the mass of the intruder, 
$\mu=(m_{\rm H}  + m_{\rm p})m_{\rm I}/(m_{\rm H} +m_{\rm p} +m_{\rm I})$ is the reduced mass of the
planet-intruder system and $a_{\rm p}$ is the semi-major axis of the planet.
If $v_{\infty} <  v_{\rm crit}$ a capture can occur \citep{2006ApJ...640.1086F}. For example, 
from equation \ref{eq:vCrit} we can calculate that if $v_{\infty} = 1.0$ km/s, 
$m_{\rm p}=M_{\rm J}$ and $a_{\rm P}=30$ au, then
 $m_{\rm I} < 30 m_{\rm J}$.

We have calculated numerically the capture cross-section for sub-stellar mass intruders
($m_{\rm I} = 3,10$ and $30 m_{\rm J}$)
in an encounter with a planetary system consisting of 
a Jupiter-mass planet orbiting a solar-mass star at 30 au using
the $N$-body code Starlab \citep{2001MNRAS.321..199P}. In Table 
\ref{tab:BDcrossSec} we list the measured cross-sections, $\, \sigma$,
scaled as the dimensionless quantity:

\begin{equation}
\sigma_{\rm scaled } = \frac{ v^{2} \sigma } { \pi a_{\rm p}^{2} }
\end{equation}
where $v$ is defined as $v=v_{\infty}/v_{\rm crit}$.

We divide the measured cross-sections into several
categories, depending on the nature of the encounter.
If a transient 3-body system forms, the encounter is 
categorised as a resonant interaction, which results
in either the planet or the intruder becoming unbound.
It is also possible that the outcome of the encounter is a prompt
exchange, where the intruder can take the place of either the
planet or the host star. Finally, if $v>1$ the total energy of the
planet-intruder system is greater than zero, and an ionisation
may occur, leaving all three objects unbound.

The measured cross-sections in Table \ref{tab:BDcrossSec} are small
if $v_{\infty} = 1$ km/s. We have also calculated the cross-sections
for $v_{\infty} = 0.5$ km/s for which the cross-sections are
an order of magnitude larger.

We have used the cross-sections in  
Table \ref{tab:BDcrossSec} to calculate the fraction of
solar-mass stars hosting a Jupiter-mass planet with semi-major
axis equal to 30 au, that are involved in resonant encounters
or direct exchanges with sub-stellar mass objects in our
reference cluster ($N=700$ stars and $r_{\rm h,initial}=0.38$ pc). 

We use the log-normal IMF of \citet{2001ApJ...554.1274C,2002ApJ...567..304C}
to extend the stellar population in our reference cluster down to one Jupiter-mass objects.
From our $N$-body simulations, we find that for encounters involving solar-mass stars
and intruder stars with $m_{\rm I} < 1 M_{\odot}$, the number of encounters that 
a solar-mass star experiences with an intruder of mass $m_{\rm I}$,
is proportional to the number of stars with mass $m_{\rm I}$ in the clusters.
We assume that this relation holds also for sub-stellar mass objects
and that the encounter rate for sub-stellar mass objects scales linearly
with $r_{\rm min}$, as is the case for stellar-mass intruders.

We use three different mass ranges of the intruders in order to account for how
the cross-sections change with mass of the intruder:  $1M_{\rm J}<m<5M_{\rm J}$
for $3M_{\rm J}$ intruders, $5M_{\rm J}<m<15M_{\rm J}$ for $10M_{\rm J}$ intruders
and $15M_{\rm J}<m<45M_{\rm J}$ for $30M_{\rm J}$ intruders. 
We list
the fractions of solar-mass stars which have been involved in
these encounters, $f_{\rm e}$, in column 5 of Table \ref{tab:BDcrossSec}.
The exact lower-mass cut-off in the IMF is not yet well constrained from observations, and
may vary between different clusters 
\citep[see, for example,][and references therein]{2009ApJ...702..805S}. 
However, we find that the fraction
of stars involved in these encounters does not change significantly if we,
for example, set the cut-off at $5 M_{\rm J}$ instead. The cause of this is
two-fold. Firstly, the capture-cross-sections increase rather strongly with
$m_{\rm I}$, and hence are smaller for intruder of mass $3M_{\rm J}$
than for the higher mass intruders. Secondly, given the IMF used, the
number of objects in the range $1M_{\rm J}<m<5M_{\rm J}$ is significantly
smaller than the number of objects with, for example, mass $15 M_{\rm J}<m<45M_{\rm J}$.

It is important to note that we have here assumed that all solar-mass stars
host planetary systems. Depending on
the velocity dispersion in the cluster, the fraction of planet-hosting stars
involved in resonant or prompt exchange encounters with sub-stellar
mass objects ranges from a few per cent  ($v_{\infty}=0.5$ km/s)
down to $10^{-3}$  ($v_{\infty}=1.0$ km/s). These fractions are less
than the fraction of four gas-giant systems from which planets are ejected
due to fly-bys with stellar-mass intruders, which is about 11 
per cent in our reference cluster (see Section 9).
The capture of a sub-stellar mass object is likely
to significantly change the orbits of planets when it occurs.
Hence, depending on the velocity dispersion in clusters, 
several per cent of planetary systems in stellar clusters could be
significantly changed by such events. We will be carrying out
more detailed simulations of rates at which sub-stellar-mass
objects are involved in encounters with planetary systems, and
how they affect the evolution of the system, in a forthcoming
paper.

\section{Changing the mass distribution of the planets}
\begin{figure}
\begin{center}
\subfigure[]{\includegraphics[scale=0.4]{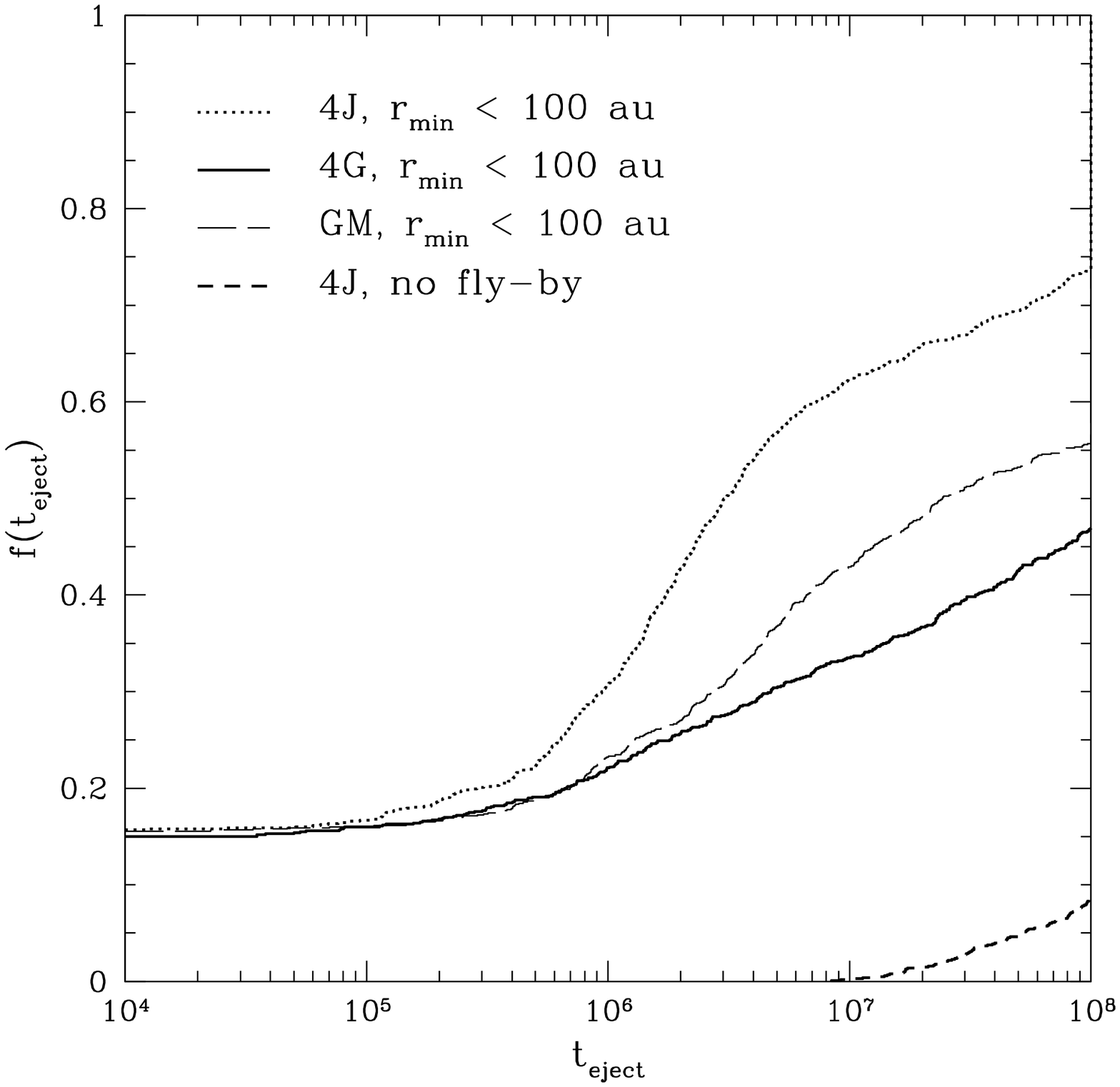}}
\caption{The evolution of the four gas giants of the solar system (4G) after a fly-by, 
the four Jupiters (4J), both if left on its own and after a fly-by and the evolution
of a systems where the mass of the gas giants in the solar systems was set as
the geometric mean (GM) between the 4G and the 4J after a fly-by. In all simulations
we varied $r_{\rm min}$ linearly between 0 and  100 au and the mass of the 
intruder star was $m_{\rm I}= 0.6 M_{\odot}$. The periastron passage 
of the intruder star occurred
at $t=0$. We plot the fraction of systems, $f(t_{\rm eject}$), from which at
least one planet has been ejected within, $t_{\rm eject}$ years after the fly-by }
\label{fig:tEject4G4JGM}
\end{center}
\end{figure}

\begin{table*}
\caption{Overview of the evolution of the four gas giants after a fly-by, where now either all four planets
have the mass of Jupiter (4J) or the planetary masses are each the geometric mean of
Jupiter and their original mass (GM).
We have divided the systems into different categories,
depending on the state of the system immediately after the fly-by. In column 1 we list the
categories and in column 2 we list the fraction of all systems which belongs to a
given category. In column 3, 4, 5 and 6 we divide the systems from each category
into four bins, depending on at what time after the fly-by the first planet was ejected from the
system, $t_{\rm eject}$. In column 3 are systems with $t_{\rm eject} < 10$ Myr, in column
4 systems with  $t_{\rm eject} < 30$ Myr, in column 5 systems with  $t_{\rm eject} < 100$ Myr
and in column 6 are systems from which no planet was ejected in $10^8$ years.
Its worth pointing out, that for systems in which a planet is ejected immediately in the fly-by,
column 4 and 5 are by definition 0, as all such planets have $t_{\rm eject} = 0$. Furthermore, the
fraction of systems which falls into the categories ``Kozai immediately after fly-by", 
``AMD unstable immediately after fly-by" and
 ´´Kozai before close encounter" are in general rather small the uncertainty in these values is
 rather large, especially in column 3-6. For example, in the case of the geometric
 mean system, only 1.6 per cent of systems, or 16 simulations, fall into the category
´´AMD unstable immediately after fly-by", with 3 systems that have $t_{\rm eject}<10^7$ years (column 3),
0 systems with $10^7<t_{\rm eject}<3\times10^7$ (column 4), 6 systems with
 $3\times10^7<t_{\rm eject}<10^8$ (column 4) and finally 7 systems which are
 AMD unstable immediately after the fly-by, but from which no planets are ejected
 within 100 Myr after the fly-by (column 6).  }

\begin{center}
\begin{tabular}{|l|c|c|c|c|c|}
\hline
&  Situation immediately  &\multicolumn{3}{l}{$t_{\rm eject}$ (Myr)} &  No planet ejection\\
& after flyby  & 10 & 30 & 100 &  after 100 Myr \\
 \hline
\multicolumn{6}{l}{The geometric mean system, $m_{\rm I} = 0.6M_{\odot}$} \\
Immediate ejection: 					&  0.153 & 1.00 & 0.00 & 0.00 & 0.00 \\
Orbit crossing immediately after fly-by:     	&  0.263 & 0.87 & 0.11 & 0.02 & 0.00 \\
Kozai immediately after fly-by:           		&  0.017 & 0.23 & 0.23 & 0.18 & 0.36 \\
AMD unstable immediately after fly-by:     &  0.016 & 0.19 & 0.00 & 0.37 & 0.44 \\
Kozai before close encounter:    		&  0.002 & 1.00 & 0.00 & 0.00 & 0.00 \\
Other (unstable):   					&  0.123 & 0.34 & 0.42 & 0.24 & 0.00 \\
Other(stable):      					&  0.426 & 0.00 & 0.00 & 0.00 & 1.00 \\
\\
\multicolumn{6}{l}{The four Jupiters, $m_{\rm I} = 0.6M_{\odot}$} \\
Immediate Ejection: 					&  0.153 & 1.00 & 0.00 & 0.00 & 0.00 \\
Orbit crossing immediately after fly-by:      &  0.263 & 1.00 & 0.00 & 0.00 & 0.00 \\
Kozai immediately after fly-by:           		&  0.017 & 0.80 & 0.13 & 0.00 & 0.07 \\
AMD unstable immediately after fly-by:     &  0.045 & 0.80 & 0.03 & 0.08 & 0.10 \\
Kozai before close encounter:       		&  0.000 &  -- 	&  --	    & --      & --       \\
Other (unstable):   					&  0.264 & 0.60 & 0.16 & 0.24 & 0.00 \\
Other(stable):      					&  0.258 & 0.00 & 0.00 & 0.00 & 1.00 \\
 \hline
\end{tabular}
\end{center}
\label{tab:GM4Jstab}
\end{table*}

In this section we consider the effects of fly-by encounters involving planetary
systems containing planets with masses different from the four gas
giants, but with the same initial orbits.
We have used two different mass-distributions: the four Jupiter
system (4J), where we set the mass of all the gas giants equal to 1 $M_{\rm J}$
and the geometric mean (GM), where we set the masses of the gas
giants, $m_{\rm i}^{*}$ as:

\begin{equation}
m_{\rm i}^{*}=\sqrt{M_{\rm J} m_{\rm i}}.
\end{equation}
where $m_{\rm i}$ is the mass of the $i^{\rm th}$ planet in the normal solar system.
We call systems in which the masses of the gas giants are all rather similar
{\sl democratic}.

We have simulated the effects of fly-bys on these systems numerically,
with $r_{\rm min}<100$ au and $m_{\rm I}=0.6 M_{\odot}$. We
performed 1000 realisations for each system, varying the initial
orientation of the intruder star, as well as the initial positions of the
planets in their respective orbits.  In 
Fig. \ref{fig:tEject4G4JGM} we plot the fraction of systems from which
at least one planet has been ejected as a function of time after the fly-by
for the 4G (solid line), the 4J (dotted line), and the 
GM (long-dashed line).
Both the 4G and GM are stable
if left alone. However, in the 4J-system, 
planet-planet scatterings are sometimes triggered without external
perturbation. We include, in Fig. \ref{fig:tEject4G4JGM}, the fraction of
systems from which at least one planet has been ejected as a function
of time in the 4J without a fly-by (short-dashed line). 

In the case of the 4J, about 8 per cent of systems  
become unstable, and eject at least one planet in $10^8$ years
without a fly-by. However, with a fly-by, the rate at which the 4J-systems
become unstable and eject planets is greatly enhanced.
About 73 per cent of the 4J-systems which have suffered
a fly-by with $r_{\rm min}<100$ au and $m_{\rm I}<0.6 M_{\odot}$
become unstable and eject at least one planet within $10^8$ 
years of the fly-by. For the GM systems, about 56 per cent of
systems eject at least one planet in $10^8$ years after a fly-by.

We have carried out the same analysis of the GM and the 4J-systems
as we did for the 4G in section \ref{sec:PredictingFBs} (see table
\ref{tab:GM4Jstab}.) As can be seen, the fraction of systems from which a planet
is ejected immediately after the fly-by, and the fraction of systems in which the
orbits of planets are found to cross, is similar to that which we find from our
simulations of the four gas giants. However, the fraction of systems that
become unstable, but do not fit into any specific category, is much larger. About
12 per cent of GM-systems and about 26 per cent of the 4J-systems
 fall into the category {\sl Other(unstable)}. Hence,
we could not predict that these systems would become unstable by analysing
the orbits of the planets immediately
after the fly-by.

\section{Varying the architecture of the planetary systems}
\begin{figure}
\begin{center}
\subfigure[]{\includegraphics[scale=0.4]{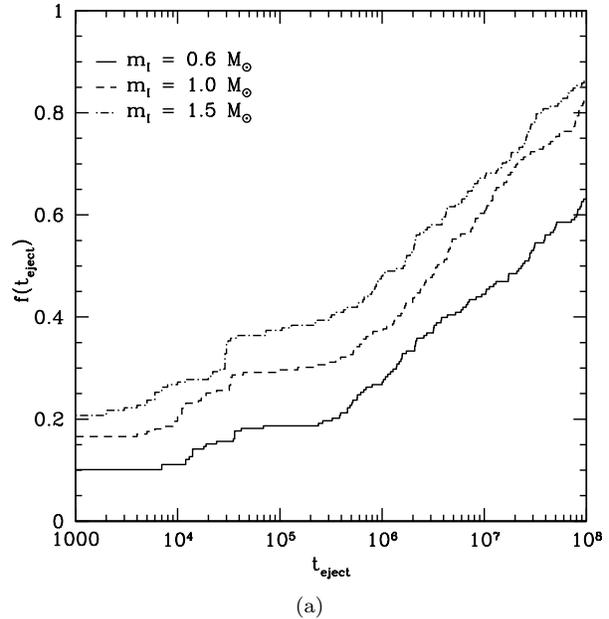}}
\caption{The evolution of the planetary system produced in run 10
of series A in \citet{1998AJ....116.1998L} after a fly-by. The properties of the
intruder star was $r_{\rm min}<100$ au 
and $m_{\rm I}= 0.6M_{\odot}$ (solid lines),  $1.0M_{\odot}$
(dashed line) and  $1.5 M_{\odot}$ (dashed-dotted line).
The periastron passage of the intruder star occurred at $t=0$.
 }
\label{fig:tEjectlev10a}
\end{center}
\end{figure}

\begin{table*}
\caption{The orbital elements of the planetary systems generated
in run 10, series A of \citet{1998AJ....116.1998L} (R10a). The columns are
mass $m$ (in Jupiter Masses, $M_{\rm J}$), semi-major axis $a$ (AU), eccentricity $e$, inclination $i$, 
argument of perihelion $\omega$, longitude of the ascending node $\Omega$, and mean anomaly $M$.}
\begin{center}
\begin{tabular}{|l|c|c|c|c|c|c|c|}
\hline
 & m ($M_{\rm J}$) & a (AU) & $e$ & $i$ & $\omega$ & $\Omega$ & M \\
 \hline
Planet 1 & 	0.880     & 4.92321     &  0.02742   &  1.49983   &  12.14303    & 108.14294   & 90.04206 \\
Planet 2 & 	0.618     & 8.34403     &  0.00684   & 1.18543    &  40.16036    & 279.69692   & 28.16437 \\
Planet 3 & 	0.140     & 11.78578   &  0.03653   &  2.79969   &  174.64686  &      0.49883  & 132.94036 \\
Planet 4 & 	0.060     & 18.69599   &  0.03278   &  3.28732   &  342.47166  & 244.48202  & 133.67521 \\
Planet 5 & 	0.041     & 32.59675   &  0.02001   &  0.98343   &    25.15793  &  287.33401  & 309.54848 \\
\hline
\end{tabular}
\end{center}
\label{tab:run10aOrbEl}
\end{table*}%

So far, we have only considered the effects of fly-bys on planetary
systems derived from our own solar system. However, as the 
population of planets around stars is likely to also include
systems with a different architecture to our solar system, we
have also simulated the effect of fly-bys on a planetary system with different
number of planets and different orbital elements than the gas giants
in the solar system.

We have used one of the systems generated
in simulations of planetary formation by \citet{1998AJ....116.1998L},
namely the system generated in run 10 of series A (henceforth called R10a). The orbital properties
of the planets in this system can be found in table \ref{tab:run10aOrbEl},
and was chosen as a system to be broadly similar to the solar system,
with gas giants on orbits between about 5 and 30 au.
This system is stable for at least $10^9$ years. It  contains five planets,
and is slightly more tightly packed than the solar system.
The masses of the planets are slightly more democratic than that
of the four gas giants of the solar system. We have performed  three
sets of fly-by simulations, with $r_{\rm min}<100$ au 
and $m_{\rm I}= 0.6M_{\odot}, 1.0M_{\odot}$ and $1.5 M_{\odot}$ 
respectively. For each set,
we have done 200 realisations, varying the initial orientation
and $r_{\rm min}$ of the intruder star. We plot the fraction of systems
from which at least one planet has been ejected in Fig. \ref{fig:tEjectlev10a}. 
As can be seen from this figure, this system is vulnerable to fly-bys.
For example, fly-bys with $m_{\rm I}=0.6 M_{\odot}$ cause 64 per
cent of these systems to undergo planet-planet scattering and eject
at least one planet in $10^8$ years, which is a number somewhat
larger than seen previously for the four gas giants in the solar system.
This is not surprising, as the system is a more packed and more democratic.

Unsurprisingly, we find that it is the two outer (and lowest mass) 
planets that are ejected most often. However, at least three planets are 
ejected in about 35 per cent of fly-bys with $m_{\rm I}=0.6 M_{\odot}$.
As seen with earlier simulations of other planetary systems, the planets
left behind have a wide range of eccentricities. The R10a system 
used
here, is thus another example of a stable planetary system which is often
destabilised via encounters in stellar clusters. The solar system is not unique;
there could be a large family of planetary architectures that follow this behaviour.

\section{Planets on very wide orbits} \label{sec:wideOffsets}
A number of  planets have been observed on very wide orbits around 
main-sequence dwarf stars by  direct imaging techniques \citep[e.g.][]{Kalas08,Marois08}. 
These imaged planets have projected separations to their respective host 
stars ranging from 25 to about 300 au. 
The host stars of these planets are A-stars with masses of $1.5-2.0 M_{\odot}$.

It is hard to form planets at such large distances from the host star in
the core-accretion scenario, as the surface density of the disc
is too low. In this section we explore how fly-bys can put planets
on very wide orbits.

\subsection{Planet-planet scatterings}

\begin{figure}
\begin{center}
\resizebox{8truecm}{!}{\includegraphics{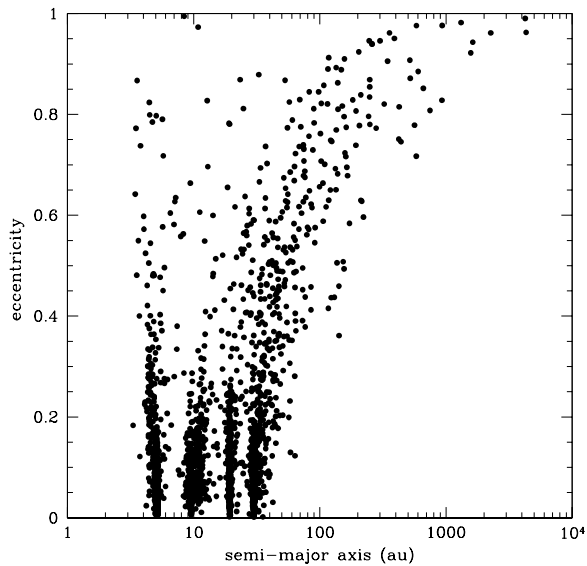}}
\caption{The semi-major axes, $a$ and eccentricities, $e$ of all planets which are bound to the
host star $10^8$ years after the fly-by in our simulations of fly-bys involving the four gas-giants
of the solar system with $r_{\rm min} < 100$ au and $m_{\rm I}=0.6 M_{\odot}$.} 
\label{fig:ae4G_10_8_years}
\end{center}
\end{figure}

\begin{figure}
\begin{center}
\resizebox{8truecm}{!}{\includegraphics{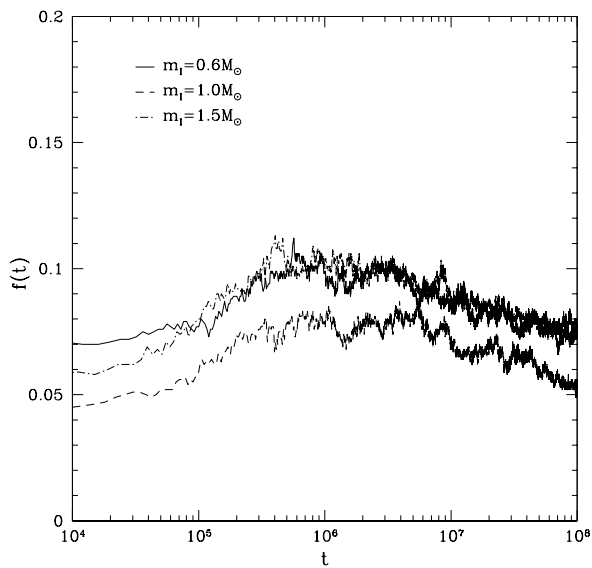}}
\caption{The fraction of systems, $f_{\rm P}$, containing planets
with semi-major axes greater than 100 au, as a function of
time, $t$, after the fly-by, in our simulations of
fly-bys involving the four gas giants of the solar system
with $r_{\rm min}<100$ au.}
\label{fig:fWide4G}
\end{center}
\end{figure}

Gas-giant planets which form at smaller orbital radii, can be put on very wide
orbits by planet-planet scattering \citep{Scharf09,Veras09}. Planet-planet scattering
often lead to the ejection of one or more planets. 
However, in general planets are not ejected in a single scattering 
event, but rather suffer a series of encounters which put them on successively
wider orbits, until they finally become unbound. This process can take several 
$10^7$ years, during which the planets can be observed on very wide orbits. 
It is important to note that while planet-planet scatterings can explain the orbits 
of single planets on wide and moderately eccentric orbits (like e.g. Formalhaut 
b),  it is very unlikely that systems like HR8799 formed this way. HR8799 consists of three 
planets which all appear to be on rather circular orbits \citep{Marois08}. 
Furthermore, it is also likely that at least two of the planets are locked in a 2:1 mean 
motion resonance \citep{2010ApJ...710.1408F}.

In Fig. \ref{fig:ae4G_10_8_years} we show a snapshot of the semi-major axes 
and eccentricities of the four gas-giant planets in the solar system, $10^8$ 
years after a fly-by with $r_{\rm min}<100$ au and $m_{\rm I}=0.6 M_{\odot}$. 
At this time, about five per cent of systems contain at least one planet with
semi-major axis larger than 100 au, which are planets that could be observed in imaging 
surveys. 

In Fig. \ref{fig:fWide4G} we plot the fraction of systems that have at least 
one planet with semi-major axis larger than 100 au as a function of time 
after a fly-by involving the 4G, with $r_{\rm min}<100$ au and $m_{\rm I}=0.6 M_{\odot}$.
As can be seen, the fraction of planets on wide orbits varies with time. 
It reach its peak value (between 0.07 and 0.11 depending on the mass of 
the intruder star) in only a few $10^5$ years
after the fly-by and then slowly decrease with time. 

From Fig.  \ref{fig:fWide4G} one can see that already at $t=10^4$ years there
are planets on wide orbits. These planets were not put on such wide orbits 
by planet-planet scattering. Instead, their wider orbits were caused by the planets
strongly interacting with the intruder star in close fly-bys. 
The fraction of systems with planets
on wider orbits is largest for fly-bys with $m_{\rm I}=0.6 M_{\odot}$. The cause is
that in these fly-bys, the perturbation on the planetary orbits is weaker than
for higher mass intruders. This decreases the fraction of systems from which
planets are ejected promptly. However, it increases the fraction of planets which are almost
ejected, and hence are put on wide orbits.

\subsection{Captured planets}
\begin{figure}
\begin{center}
\includegraphics[scale=0.4]{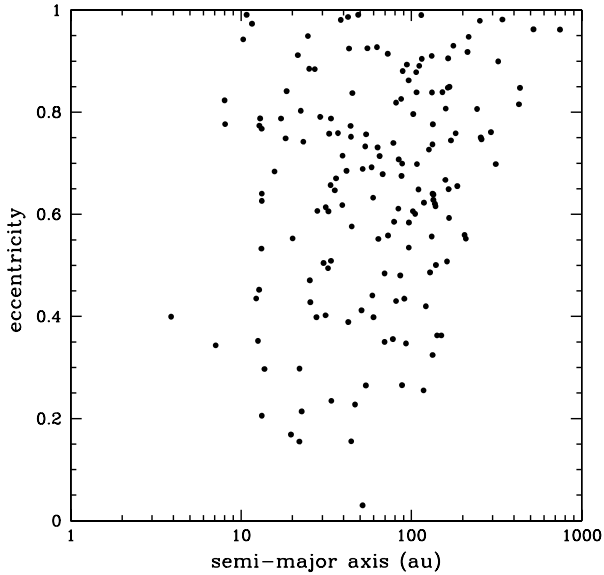}
\caption{The semi-major axes, $a$, and eccentricities, $e$, of planets that were captured by the intruder star
in fly-bys with $r_{\rm min}<100$ au, involving the four gas-giant planets of the Solar system and a intruder star with mass
$1.5 M_{\odot}$. We here only include planets in single-planetary systems, i.e. in which only one
planet was captured by the intruder star. Most of the planets are on rather wide orbits, with about
40 per cent having $a>100$ au. }
\label{fig:aeSingleInt}
\end{center}
\end{figure}

In tight fly-bys, where the minimum separation between the host star 
and the intruder is only a factor two or three larger than the semi-major 
axis of the outermost planet, one or more planets may become bound to 
the intruder star during the encounter. In Fig \ref{fig:nBoundrMin} we plot the probability that 
one planet is captured by the intruder star in encounters involving the four gas giants of the 
solar system, where $r_{\rm min}<100$ au (thick lines). As an example, in our reference cluster, the fraction
of stars in the mass-range $1.3<m<1.7 M_{\odot}$ that would capture one planet in fly-bys
is $0.08 \times f_{\rm p}$, where $f_{\rm p}$ is the fraction of stars in the cluster hosting
planetary systems similar to the 4G.

Planets which are captured during a fly-by are in general left on rather wide and 
moderately eccentric orbits. In Fig. \ref{fig:aeSingleInt} we plot the 
semi-major axes and eccentricities of the planets  captured  by a $1.5 M_{\odot}$ intruder.  
About 40 per cent have semi-major axes larger than 100 au. If a
fly-by occurs when the planets are newly formed, and hence are still contracting, the 
luminosity of the planets, combined with their large separations to the host star, should
make them detectable in imaging surveys.

\section{Properties of planetary systems produced in stellar clusters} \label{sec:plSysProducedInStellarClusters}
\begin{figure}
\begin{center}
\includegraphics[scale=0.4]{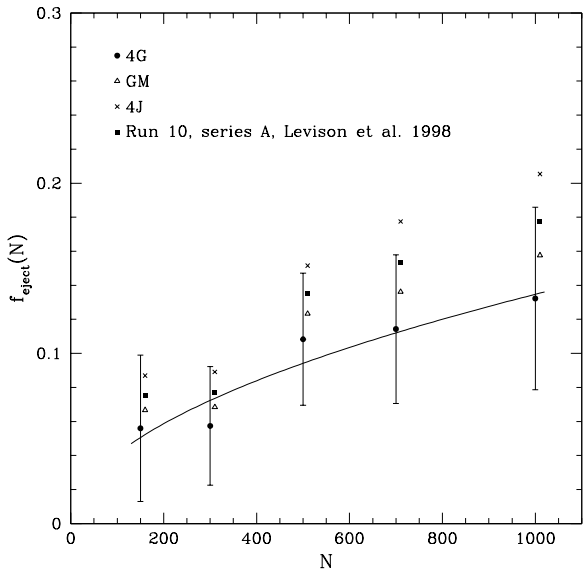}
\caption{The fraction of planet-hosting systems from which 
at least one planet is ejected in $10^8$ years due to fly-bys.
The fractions were calculated assuming a single value of the
intruder star mass of $m_{\rm I}=0.6 M_{\odot}$, with $r_{\rm min}<100$ au.
The error bars are the standard deviation from the mean of $f_{\rm eject}$
as measured for the 4G. In the figure we have, for clarity, offset $N$ slightly to the
right for all systems except the 4G. }
\end{center}
\label{fig:MeasuredFracs}
\end{figure}

\begin{table}
\caption{The fraction of planetary systems of the four gas giants (4G), four
Jupiters (4J), the geometric mean of the two (GM) and the planetary system produced in run 10
of series A in \citet{1998AJ....116.1998L} (R10a)  from which planets
are immediately ejected, $f_{\rm ion}$, captured by the intruder star, $f_{\rm cap}$, and the total
fraction of systems from which planets are ejected due to fly-bys. $f_{\rm eject}$.
The fractions listed were calculated for clusters with an initial 
half-mass radius of 0.38 pc, and only for encounters with $m_{\rm I}=0.6M_{\odot}$. 
The calculated fractions assume that all solar-mass stars host a planetary
system of the given type. In order to calculate the fraction of all
stars affected, the values listed below
must be multiplied by $f_{\rm p}$, the fraction of
stars which hosts planetary systems. }

\begin{center}
\begin{tabular}{llccc}
\hline
 & N & $f_{\rm ion}$  & $f_{\rm cap}$  & $f_{\rm eject}$  \\ 
\hline
4G 		& 150 	&   	$1.3 \times 10^{-2}$	&  $7.2 \times 10^{-4} $	&  $5.6 \times 10^{-2} $ 	\\
     		& 300 	&   	$1.4 \times 10^{-2} $	&  $7.4 \times 10^{-4} $	&  $5.7 \times 10^{-2} $	\\
   		& 500 	&   	$1.9 \times 10^{-2} $	&  $1.0 \times 10^{-3} $	&  $1.1 \times 10^{-1} $	\\ 
		& 700 	&   	$2.7 \times 10^{-2} $	&  $1.5 \times 10^{-3} $	&  $1.1 \times 10^{-1} $	\\ 
   		&1000	&  	$3.2 \times 10^{-2} $	&  $1.7 \times 10^{-3} $	&  $1.3 \times 10^{-1} $	\\
															 				\\                                                                                            
GM 		& 150 	&    	$1.3 \times 10^{-2} $	&  $7.2 \times 10^{-4} $	&  $6.7 \times 10^{-2} $	\\
   		& 300 	&      	$1.4 \times 10^{-2} $	&  $7.4 \times 10^{-4} $	&  $6.8 \times 10^{-2} $	\\ 
 		& 500 	&     	$1.9 \times 10^{-2} $	&  $1.0 \times 10^{-3} $	&  $1.2 \times 10^{-1} $	\\ 
   		& 700 	&     	$2.7 \times 10^{-2} $	&  $1.5 \times 10^{-3} $	&  $1.4 \times 10^{-1} $	\\
   		&1000 	&    	$3.2 \times 10^{-2} $	&  $1.7 \times 10^{-3} $	&  $1.6 \times 10^{-1} $	\\ 
	 																		\\                                                                                            
4J 		& 150 	&    	$1.3 \times 10^{-2} $	&  $7.2 \times 10^{-4} $	&  $8.7 \times 10^{-2} $	\\
   		& 300 	&      	$1.4 \times 10^{-2} $	&  $7.4 \times 10^{-4} $	&  $8.9 \times 10^{-2} $	\\
   		& 500 	&     	$1.9 \times 10^{-2} $	&  $1.0 \times 10^{-3} $	&  $1.5 \times 10^{-1} $	\\
   		& 700 	&     	$2.7 \times 10^{-2} $	&  $1.5 \times 10^{-3} $	&  $1.8 \times 10^{-1} $	\\
   		&1000 	&    	$3.2 \times 10^{-2} $	&  $1.7 \times 10^{-3} $	&  $2.1 \times 10^{-1} $	\\
																			\\                                                                                             
R10a	& 150 	&    	$7.2 \times 10^{-3} $	&  $7.2 \times 10^{-3} $	&  $7.5 \times 10^{-2} $	\\
   		& 300 	&      	$7.4 \times 10^{-3} $	&  $7.4 \times 10^{-3} $	&  $7.7 \times 10^{-2} $	\\
   		& 500 	&     	$1.0 \times 10^{-2} $	&  $1.0 \times 10^{-2} $	&  $1.4 \times 10^{-1} $	\\
   		& 700 	&     	$1.5 \times 10^{-2} $	&  $1.5 \times 10^{-2} $	&  $1.5 \times 10^{-1} $	\\
   		& 1000 	&    	$1.7 \times 10^{-2} $	&  $1.7 \times 10^{-2} $	&  $1.8 \times 10^{-1} $	\\
 \hline
\end{tabular}
\end{center}
\label{tab:fracPlSys}
\end{table}

In this section we describe how the rate at which planets are ejected from solar-system
like planetary systems vary with the properties of the stellar cluster. 
Of course, what we are interested in here, is to understand what
fraction of planetary systems are significantly affected by fly-bys.  In such
systems, planet-planet scattering often leads to the ejection of one or more
planets. Hence, the fraction of systems which eject planets is a good 
measure on the fraction of systems significantly changed by fly-bys.

 In the context of the extrasolar planet population, we are
interested in  the properties of the planets that are left orbiting
stars after fly-bys. We describe some of these properties in the second part of the section.

\subsection{Fraction of planetary systems damaged by fly-bys in stellar clusters} 
We calculate the fraction of planetary systems from which planets are
ejected, both immediately and in total, as well as the fraction of systems from
which one or more planets is captured by the intruder in fly-bys as follows. 
First, we measure the number of stars
in our stellar cluster simulations that  have undergone fly-bys with
$r_{\rm min}<100$ au, $N_{\rm fb}$. We have included the effects of multiple fly-bys in the calculation in
a simplified way, by assuming that
a star which has had $N_{\rm fb}$ fly-bys, is equivalent to $N_{\rm fb}$  stars
having had one fly-by each. By doing this we neglect the combined effect of multiple
fly-bys on one planetary system.
Once we know $N_{\rm fb}$, we can multiply this number by the
fraction of  fly-bys having $r_{\rm min}<100$ au that lead to ionisation,
capture and/or the
ejection of at least one planet in $10^8$ years after the fly-by.
Here, we only include fly-bys with intruder stars of mass
$m_{\rm I}=0.6M_{\odot}$. Having obtained the number of planetary
systems, we then divide this by the
number of stars hosting planetary systems in the cluster, to obtain the fraction
of planetary systems which involved in ionising encounters, $f_{\rm ion}$,
the fraction of system involved in fly-bys which lead to the capture of
at least one planet, $f_{\rm cap}$, as well as the fraction of planetary systems that
eject planets in $10^8$ years after fly-bys, $f_{\rm eject}$.

In Table \ref{tab:fracPlSys} we list $f_{\rm ion}$, $f_{\rm cap}$ and $f_{\rm eject}$ in clusters with 
$r_{\rm h,i}=0.38$ pc for the 4G, GM,
4J and R10a systems.

The values of $f_{\rm ion}$, $f_{\rm cap}$ and $f_{\rm eject}$ listed in Table \ref{tab:fracPlSys} are lower limits.
When calculating them we used the fraction of fly-bys causing ejections and/or capture of planets
as calculated for $m_{\rm I}=0.6 M_{\odot}$.  
However, as can be seen in, for example, Table 1, fly-bys with higher mass
stellar intruders are more damaging to planetary systems. We can, for the 4G
system, include the increased damage done by fly-bys with higher mass as
follows.
We divide the encounters into three mass-bins
($0.2M_{\odot}<m_{\rm I}<0.8M_{\odot}, \, 0.8M_{\odot}<m_{\rm I}<1.2M_{\odot}$
and $\, 1.2M_{\odot}<m_{\rm I}<5M_{\odot}$) and calculate what fraction of
solar-mass stars have had an encounter within $r_{\rm min}<100$ au for each
mass bin. We then multiply this number with the fraction of such fly-bys which
cause the ejection of at least one planet in $10^8$ years after the fly-by. We 
find that $f_{\rm eject}$ for the 4G is then increased, from 0.11 to 0.14 ín our
reference cluster.

In Fig. 13  we plot the fraction of systems, $\, f_{\rm eject}$, from which planets
are ejected in $10^8$ years after a fly-by, as a function of the number of 
stars, $N$, in the clusters. We include the standard deviation of $\, f_{\rm eject}$
as calculated for the 4G system. This uncertainty is the standard deviation from
the mean of the fraction of stars which undergo encounters in our cluster
simulations.

The solid line in Fig. 13 is a least-squares fit to $\, f_{\rm eject}$,
as measured for the 4G system, where we assume that:

\begin{equation}
f_{\rm eject} \propto N^{\gamma}.
\end{equation}
 The best fit gives $\gamma = 0.52 \pm 0.11$.

The fraction of stars, $f_{\rm enc} $, which encounter another star within a distance $r_{\rm min}$, 
during the lifetime of a stellar cluster, $\tau$, can be estimated as \citep{2003gmbp.book.....H}:

\begin{equation}
f_{\rm enc} = \int _0^{\tau} n \sigma v dt
\end{equation}
where $n$ is the number density, $\sigma$ is the cross-section of the encounter and
$v$ is the velocity dispersion.

The velocity dispersion in a cluster depends on its radius, $r_{\rm h}$ and mass, $M$
\citep{2003gmbp.book.....H}:

\begin{equation}
v^2 \propto \frac{M}{r_{\rm h}} \propto \frac{N}{r_{\rm h}}
\end{equation}
where we in the last step use the relation $M=\bar{m} \times N$, with $\bar{m}$
the average mass of stars in the cluster and $N$ the number of stars. 

However, all the clusters in our simulations have roughly the same size during
most of their lifetime ($r_{\rm h} \sim 2$ pc), and hence we can write:

\begin{equation}
v^2 \propto N
\end{equation}

The cross-section, $\sigma$ goes as $\sigma \propto v^{-2}$ for a given value of $r_{\rm min}$
when dominated by gravitational focussing (see Equation \ref{eq:CrossSection}),
 and hence it scales with $N$ as:

\begin{equation}
\sigma \propto \frac{1}{v^2} \propto \frac{1}{N}
\end{equation}
while the number density, $n$ scales linearly with the number of stars,
$n \propto N$, as $r_{\rm h}$ is constant. 
 
The lifetimes of clusters, $\tau$, increase with the number of stars, $N$. 
In all the clusters, 50 per cent of fly-bys occur in the first $10^7$ years,
and 95 per cent of fly-bys in the first $10^8$ years. As an order of
magnitude estimate, we can thus write:

\begin{equation}
f_{\rm enc} \sim \int _0^{\tau_0} n \sigma v dt
\end{equation}
where $\tau_0=10^8$, as the lifetime of all the clusters we have simulated 
are longer than $10^8$ years. Combining
these equations we thus find:

 \begin{equation}
f_{\rm enc} \propto N^{1/2}
\end{equation}

We assume that the fraction of fly-bys within a given value of $r_{\rm min}$,
which damage planetary systems, does not change with $N$. Then we can
write:

\begin{equation} \label{eq:fEjectWithN}
f_{\rm eject} \propto N^{1/2}
\end{equation}

This result agrees well with our $N$-body simulations.
That $f_{\rm eject}$ depends only weakly on $N$ is particularly
interesting in terms of the the cumulative probability, $P(N)$, that a 
given star forms in a cluster
of size $N$. In the solar neighbourhood, observational surveys suggest that 
\citep[see][and references therein]{2010arXiv1001.5444A}:

\begin{equation} \label{eq:clProb}
P(N) \propto \ln{N}
\end{equation}
for $N=100-2000$. This means that all of our clusters simulated
will contribute roughly equally to the stellar population.
Given that $f_{\rm eject}$
varies only weakly with $N$, low-$N$ clusters can thus play
an important role in populating the solar neighbourhood with
planetary systems which have undergone planet-planet
scattering caused by fly-bys.

\subsection{Properties of the planet population}

\begin{figure}
\begin{center}
\resizebox{8truecm}{!}{\includegraphics{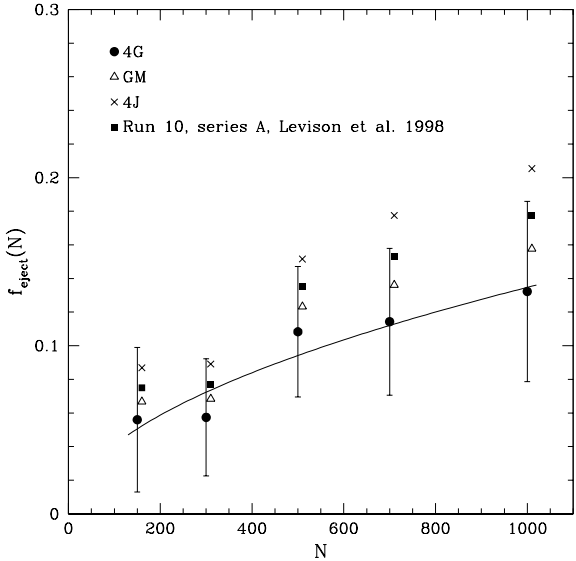}}
\caption{The eccentricity distribution of planets with semi-major
axisin the range $0.1<a<4.5$ au, both observed (detected using radial velocities only, 
long-dashed line), and from our simulations of fly-bys involving the 4G (solid
line), GM (short-dashed line) and 4J (dotted line).}
\label{fig:eccDists}
\end{center}
\end{figure}

In terms of understanding how fly-bys can affect a given population
of planetary systems, we must look at the properties of these
planetary systems long after the fly-by. In this section we
discuss the eccentricities and semi-major axes of the planetary
systems left around the host star $10^8$ years after a fly-by.

\subsubsection{Eccentricities}
In Fig. \ref{fig:eccDists} we plot the
eccentricity distribution of the observed planets detected using
radial velocities (long-dashed line). We also plot the eccentricity
distribution of 4G (solid line), GM (long-dashed line) and 4J
(dotted line) planetary systems as well as the system of run 10,
series A from \citep{1998AJ....116.1998L} (dash-dotted line). 
The planetary systems are ``observed" $10^8$ 
years after the fly-by, i.e. at the end of our simulations. 
When producing these distributions, we have included all planetary
systems in our fly-by simulations (i.e. also those from which no
planets were ejected in $10^8$ years). However, we only
include planets on orbits with a semi-major axis less than 4.5 au,
which is roughly the current detection limit in radial velocity surveys. 
Furthermore, we have limited the sample to planets on orbits wider
than 0.1 au. Planets on tighter orbits than this will be tidally
circularised, leading to a pile up of planets with $e=0$ in the 
observed sample, which is not representative of the mechanism
which excite the eccentricities. 
All of the planetary systems which we have simulated here,
have gas giants on orbits with semi-major axis 
larger than 4.5 au if no scattering has occurred.

As can be seen, the eccentricity distribution of the
4G post-scattering systems is rather similar to the
observed eccentricity distribution. Using the 
Kolmogorov-Smirnov statistic, which determines
how different two sets of data are, we find that the 
two samples are consistent with originating from 
the same underlying population ($P_{\rm KS}=0.61$).

We caution the reader that while the 4G eccentricity distribution
appears to fit the observed planets rather well, it is 
important to note that we have excluded any contribution
from unperturbed planetary systems in our data.
In order to explain the observed semi-major axis distribution,
many planetary  systems must contain planets with semi-major axes
less than 4.5 au initially (i.e. the planets 
must have migrated there in a proto-planetary disc). The
observed population thus contain such unperturbed
systems with planets on rather circular orbits while the
4G population in Fig. \ref{fig:eccDists} does not.

It is interesting to note that eccentricity distribution of planets 
in the post-fly-by 4G systems (Fig.  \ref{fig:eccDists})
differs from those which we find when simulating these planetary systems
in binaries, where hierarchical systems, like the 4G, produced much
fewer eccentric systems than the observed distribution \citep{2009MNRAS.394L..26M}.
The cause of this is that in fly-bys, the eccentricity of the inner planet
is not only excited by planet-planet scatterings, as is the case in
binaries where the Kozai mechanism operate, but also by direct
interaction with the intruder star in very tight fly-bys.

\subsubsection{Semi-major axes}
In planet-planet scattering, the total energy of the
planetary system is conserved. We can estimate the 
smallest semi-major axis which a planet can have after
planet-planet scattering. The total binding energy
in a planetary system, $E_{\rm bind}$, depends only on the mass
of the planets, $m_{\rm p,i}$, their semi-major
axes, $a_{\rm p,i}$ and the mass of the host star, $M_{\rm H}$:

\begin{equation} \label{eq:ePlSys}
E_{\rm bind}= - \sum_{i=1}^{n_{\rm pl}}  \frac{G M_{\rm H} m_{\rm p,i}}{2a_{\rm p,i}} 
\end{equation}
where  $n_{\rm pl}$ is the number of planets in the
system.

Assuming that all but one planet is ejected in planet-planet
scatterings, the semi-major axis of the remaining planet, $a_{\rm f}$,
is: 

\begin{equation} \label{eq:aFin}
a_{\rm f} = \frac{m_{\rm f}}{\sum_{i=1}^{n_{\rm pl}} m_{\rm p,i}/a_{\rm p,i}  }
\end{equation}
where $m_{\rm f}$ is the mass of the remaining planet. Here we
have assumed that the kinetic energy of the ejected planets
is negligible.

In, for example, the 4G system, where most often Jupiter is the 
remaining planet after scattering, we find $a_{\rm f}= 4.4$ au,
assuming that the planets were on their original orbits at
the onset of scattering (i.e. the fly-by only changed the eccentricities
of the planets). In more democratic systems, the semi-major axes
can be decreased by a larger factor. In, for example,
4J equation (\ref{eq:aFin}) gives $a_{\rm f} = 2.6$
au. However, many extrasolar planets are found on orbits
much tighter than 2.6 au. 
Hence, although the eccentricities of the 
post-fly-by systems in Fig. \ref{fig:eccDists} match the 
observed distribution, their semi-major axes distribution does not. 
The observed planets on tighter orbits can, however, be produced 
by planet-planet scattering if the semi-major axes of the gas giants 
are significantly smaller than in the solar system at the onset of 
planet-planet scattering. 
We are therefore lead to conclude that migration must be working
to bring gas giants in to distances of about 0.5 - 2 au before the system becomes
unstable on its own or via the perturbations of passing stars or stellar binary companions.

\section{Discussion} \label{sec:Discussion}
\subsection{Fly-bys involving binary intruders}
When calculating the rate at which planets are ejected from planetary 
systems one should remember that the intruder may be a binary system. 
Such encounters may be more damaging to planetary
systems than fly-bys of single stars \citep[see also][]{2001Icar..150..151A}. Currently, we have
included such encounters by setting the mass of the
intruder equal to the sum of the masses of the binary components.
However, by doing this, we neglect the possibility that these encounters
may lead to the formation of a transient stellar 3-body system, which 
could potentially be very damaging to any planetary system around the
single star. 

From our $N$-body simulations, we find that in about 1 in 15 fly-bys
which involve solar-mass stars, the intruding star is a binary
system. As one third of stars in our cluster simulations are in 
binaries, one might expect the rate to be significantly higher. 
However, binaries are destroyed in encounters between two
binaries, thus the population of binaries is depleted over time
\citep[see, for example,][]{1992ApJ...389..527H}.

Given the the binary population is really rather small in our clusters
for most of the time, the contribution made to the fly-by encounters
is small. If further, we assume that all encounters involving a binary
lead to the ejection of at least one planet then we find that  
$f_{\rm eject}$ for the 4G, as measured at $10^8$ years, 
increases from 0.11 to 0.12. 

However,
if the primordial binary fraction in the clusters were higher, which is suggested
by observations \citep[e.g.][]{duq91}, the rate of fly-bys involving binaries would
also be much larger. This in turn would cause a larger fraction of planetary systems
to be strongly affected by fly-bys in clusters. We will present the results of simulations
of clusters with a higher primordial binary fraction in a forthcoming paper.

\subsection{The effects of fly-bys on protoplanetary discs}
In this paper we only consider the effects of fly-bys after the
phase of giant-planet formation is complete. However, an important aspect of 
fly-bys is also how they affect the planet-formation process.

A fly-by on a protoplanetary disc can, for example,
truncate the disc. \citet{2001Icar..153..416K}
find, using both analytical calculations and numerical
simulations, that discs can be truncated at a radius $r_{\rm d}=r_{\rm min}/3$
in encounters with other stars, thereby halting planet 
formation at radius larger than $r_{\rm d}$. Hence, if, for example,
a star has a fly-by with $r_{\rm min} \approx 100$ au, the disc will
be truncated at a radius of about 30 au, the semi-major axis
of Neptune. 

Fly-bys can also affect the properties of planets which form through core-accretion.
Simulations of parabolic fly-bys of discs with on-going planet
formation, suggest that fly-bys may cause planets
to migrate outwards, and increase the final mass of the planets \citep{2009A&A...505..873F}.

It is also possible that fly-bys can trigger 
instabilities in protoplanetary discs, inducing planet
formation via gravitational instability. 
Simulations by  \citet{2009MNRAS.400.2022F}
suggest that this is, however, not the case and 
that fly-bys instead tend to halt
any fragmentation in the disc, thereby decreasing the rate at 
which planets form.

The high number density of stars in stellar clusters means that radiation from nearby
stars can affect the planet formation process via photo-evaporation. Photo-evaporation
of discs decreases, or halts, planet formation. The rate at
which this occurs in embedded clusters was studied by \citet{2004ApJ...611..360A,
2006ApJ...641..504A}.
They find that in 10 million years, photo-evaporation in a cluster containing about 1000 stars
can truncate protoplanetary discs
down to a radius, $r_{\rm d}$, of $r_{\rm d} = 36 \, \mbox{au} \, (M_{\rm H}/M_{\odot}$),
where $M_{\rm H}$ is the mass of the planet host star. Thus, in such a cluster,
photo-evaporation could affect giant planet formation, in particular around
stars of somewhat lower mass than the Sun.

\subsection{Planetary systems in primordial binaries}
In our simulations of stellar clusters \citep{2007MNRAS.378.1207M}, one third of 
stars are in primordial binaries. Observations \citep[e.g. ][]{duq91} suggest that
the actual binary fraction may be much higher than this, as about 50 per cent
of stars in the field are observed to be part of a binary system. Increasing the
binary fraction in our simulations would increase the rate of exchange encounters,
thereby increasing the fraction of planetary systems which spent time in
a binary system.

In this paper we only consider the effects of encounters on planetary 
systems around initially single stars. It is however
quite possible that planetary systems also form in primordial binary systems
\citep[e.g][]{2002A&A...396..219B,2004A&A...427.1097T,2006Icar..183..193T}.
While the inclination between the binary companion and the planets
in such a system may initially not be large enough for the Kozai
mechanism to operate, fly-bys can change the orientation of the companion star.
\citet{2009MNRAS.397.1041P} simulated the evolution of young embedded clusters
for 10 Myr, with all stars in primordial binaries (i.e. the binary
fraction equalled one). They found that 20 per cent of binary systems 
(which were initially all assumed to be co-planar) 
attained an inclination high enough for the Kozai Mechanism 
to operate. This fraction may become even higher in long-lived clusters,
such as those studied here. We will re-visit this issue in a forthcoming paper.

\subsection{Fly-bys involving planetary systems with planets on tight orbits}
As mentioned in Section 9.2, fly-bys involving planetary systems with similar
properties to the solar system (i.e. the systems discussed in this paper),
cannot reproduce the semi-major axes distribution of the observed planets, $f_{\rm obs} (a)$.
To explain the shape of $f_{\rm obs} (a)$, the primordial population of planetary
systems (i.e. the population of planets before planet-planet scattering), must contain
systems with planet on orbits significantly inside the ice line. Hence, many planets must have
undergone disc migration.

In the current study we have not simulated the evolution of such tighter systems
involved in stellar fly-bys. As such, we are not currently able to quantify how the
rate at which such systems are significantly changed in stellar clusters would
differ with respect to solar-system-like planetary systems. However, we know that two competing
effects will play an important role:

\begin{enumerate}
\item {\sl Increasing $a_{\rm outer}/r_{\rm min}$:} For a given fly-by, the ratio between the
semi-major axis of the outermost planet, $a_{\rm outer}$, and $r_{\rm min}$ determines
the magnitude of the perturbation on the planetary orbits. As such, one might expect that the
instant effect of fly-bys involving tight planetary systems will be smaller than for fly-bys
involving solar-system-like planetary
systems.

\item {\sl The stability of tight planetary systems:} A system which has undergone disc
migration may be more tightly packed than a solar-system-like planetary systems. As 
such, the perturbation needed in such a system to trigger planet-planet scattering might
be considerably smaller. Their stability may also crucially depend on two or more planets being in a mean-motion
resonance with each other \citep[see, for example,][]{2004ApJ...611..494B}. Even a rather small perturbation,
which breaks the mean-motion resonance, might then be enough to trigger planet-planet 
scattering.
\end{enumerate}

Hence, it is not clear how the effects of fly-bys, and how they affect the properties of the primordial
population of planets, will change if many planetary systems contain planets which have
migrated onto much tighter orbits. We will study this further in a forthcoming paper.

\subsection{The formation of extrasolar planets}
\begin{figure*}
\begin{center}
\resizebox{!}{12truecm}{\includegraphics{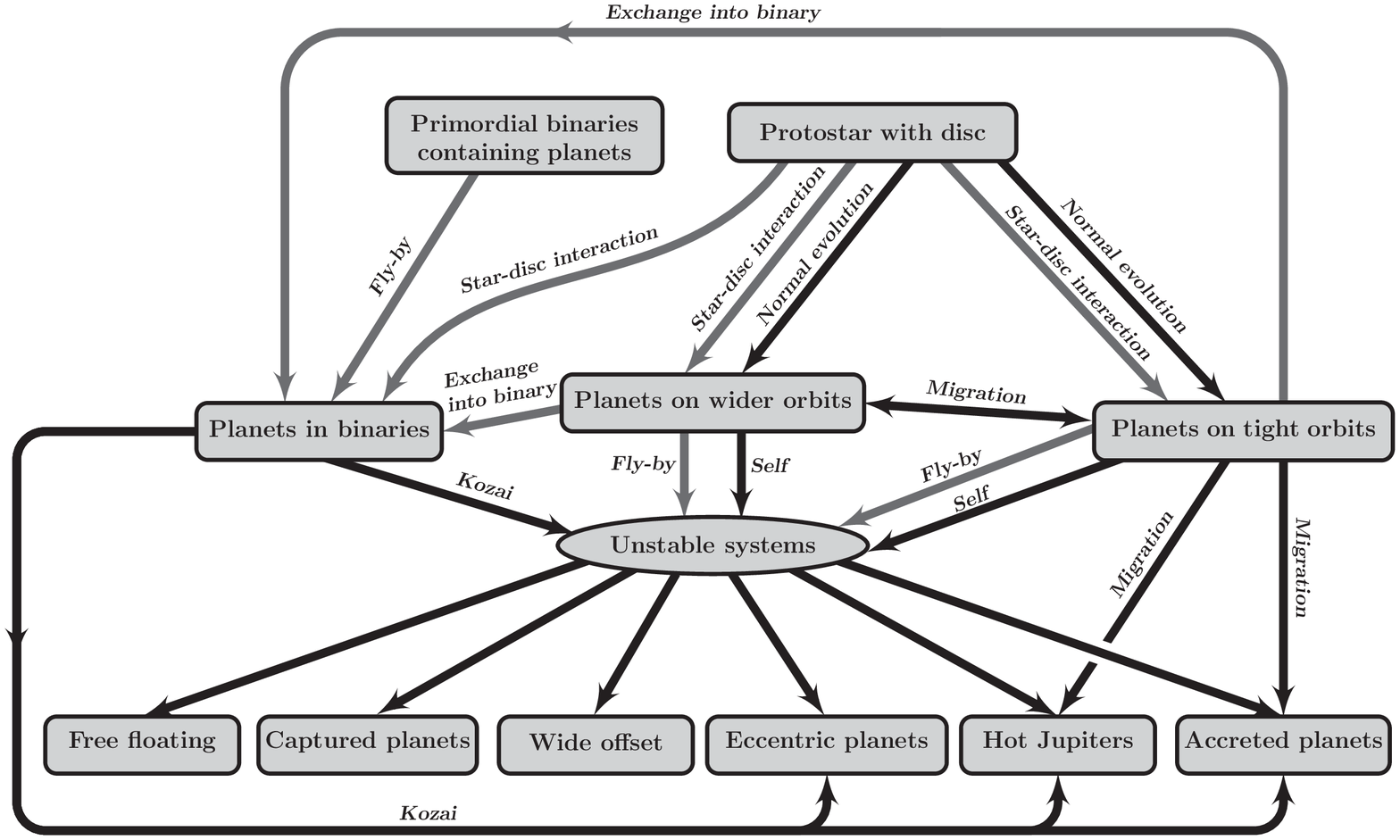}}
\caption{Overview of the evolution of planetary systems. Grey arrows 
identify interactions with other stars, most of which occur in young stellar 
clusters. Examples of such are fly-bys and exchange encounters 
involving binary systems. Planetary systems similar to our own solar system
belong to the box labelled ``Planets on wider orbits".}
\label{fig:flowChart}
\end{center}
\end{figure*}

In Fig. \ref{fig:flowChart} we show an overview of the possible 
evolutionary scenarios of planetary systems. The boxes show different
types of systems, and the arrows identify how these may evolve. 
Planetary systems resembling the solar system, like those 
simulated in this paper, belong to the category ``Planets
on wider orbits".

Black arrows
identify an evolutionary scenario occurring without influence from any external bodies, while
grey arrows identify evolutionary scenarios triggered by 
perturbations from other stars. 

For example, a ``Protostar with disc" can evolve into a system resembling
the solar system, either through normal evolution (e.g core-accretion) or
through fly-by induced fragmentation of the disc (e.g. gravitational instability).

Such systems may remain stable, like the solar system, or dynamically evolve.
If the system is long-term stable, as is the case for four gas giants, a fly-by or exchange
encounter can turn it into an ``Unstable system". If the planetary system is
initially not stable, as is the case of the four Jupiters, the systems will pass
into the box labelled unstable systems. 

Unstable systems undergo planet-planet scattering, producing the types of
extrasolar planetary systems which we observe today in radial velocity, transit and
imaging surveys.

One interesting possibility is that planetary systems which undergo
exchange encounters will sometimes find themselves in extremely
inclined binaries. If so, the Kozai mechanism can excite the eccentricities
of planets to values close to unity. Planets
will then have a very small periastron distances, causing them to tidally
interact with the host star. The tidal interaction causes the
planets' orbits to be circularised, greatly decreasing their semi-major axes. This 
behaviour, often called Kozai migration, can lead to the formation
of hot Jupiters from planets on initially much wider orbits 
\citep[see, for example,][]{2007ApJ...669.1298F}.
A key point here is that Kozai migration will often leave the
hot Jupiters significantly inclined with respect to the rotational axes of the host
stars. For transiting hot Jupiters, it is possible to measure the projected
inclinations of the planets' orbits with respect to the stars' rotational
axes, the so-called Rossiter-McLaughlin effect. Recently, \citet{triaud10}
showed that the distribution of inclinations for a large set
of observed hot Jupiters follows that expected if they were all made through
Kozai migration, and is inconsistent with formation via e.g.
type II disc migration.

Planet-planet scatterings will ordinarily place
planets on transient wide orbits (i.e. with semi-major axes greater than
100 au) in both the intrinsically unstable systems and in those systems
where instabilities are induced by encounters. 
Planetary systems which undergo
scatterings are in general only a few 100 million years old. Hence,
if planet-planet scattering is responsible for the eccentricities that we measure in 
the observed exoplanet systems,
planets on wide orbits will be observed in imaging
surveys of nearby solar-type stars \citep[e.g.][]{2010ApJ...714.1551H}.

\subsection{Observational signatures of fly-bys}
As shown in this paper, the principal effect of fly-bys is that they may trigger
planet-planet scattering in otherwise stable planetary systems. However, the
post-scattering systems cannot easily be distinguished from systems in which 
the scattering was triggered because the systems were intrinsically unstable
\citep[e.g.][]{2008Icar..193..475M,2008ApJ...686..603J,2008ApJ...686..580C}.

One important aspect of fly-bys is, however, that only a fraction of stars
suffer fly-bys. Hence, a large fraction of stars are singletons, and any
planetary systems around them have not been perturbed by other
stars. Hence, we predict that once planets can be detected in
the semi-major axes region between 5 and 
30 au around other stars many 
systems with similar properties to the solar system should be observed,
if most systems start out with similar properties to the solar system.

Furthermore, as stellar encounters occur throughout the entire life-times
of stellar clusters, we would expect that the fraction of planetary systems
which are perturbed increase with cluster age. Hence, the properties of the
population of planets in stellar clusters will depend on the age of the cluster.

Another important observational signature of planet-planet scattering, whether
it is triggered by interactions with other stars or not, is,
as is described both in this paper and in \citet{Scharf09,Veras09}, that it will
produce planets on very wide orbits. Such planets must be observed in
future imaging surveys, if planet-planet scattering is responsible for
the observed eccentricities among extrasolar planets.

\section{Summary}
We have numerically simulated fly-bys involving planetary systems
resembling our own solar system. Our motivation was to understand
how a population of planetary systems evolve in stellar clusters, the birth
environment of a large fraction of all stars. We have combined the fly-by
simulations with our previous simulations of young stellar clusters, to
calculate the effect on the population of planetary systems in the field
today. The main results of the paper are:

\begin{enumerate}
\item  Close fly-bys can lead to the immediate ejection of one or 
more planets from the host system. Such planets are 
either left unbound (ionisation), or are captured by the 
intruder star (exchange). Captured planets
are often left on wide and moderately 
eccentric orbits, leaving them detectable in imaging surveys. 
Planets which remain bound to the host star after a close
fly-by are most often on significantly more eccentric orbits. 
In a moderately-sized stellar cluster ($N=700$ stars, 
$r_{\rm h}=0.38$ pc) about 3 per cent of solar-system-like 
planetary systems will be involved in fly-bys leading to 
ionisation and/or capture of planets.

\item Wider fly-bys, which do not immediately remove planets from the host star, may still cause 
the ejection of planets on longer time scales by planet-planet 
scattering. The time between the fly-by and the onset of planet-planet scattering can be very long.
For the four gas giants of the solar system we find that we can accurately predict if the system will
suffer planet-planet scattering within $10^8$ years after a fly-by (e.g. Table 1), while the same is 
not easily done for the system where the masses of the planets have all been set to
that of the planet Jupiter (4J) (e.g. Table 3). Planet-planet scattering significantly increases the fraction
of planetary systems from which planets are ejected due to fly-bys. In, for example, the case of the 4G planetary 
systems and encounters with $r_{\rm min}<100$ au and $m_{\rm I}=0.6M_{\odot}$ the fraction
of systems which suffer ejections is increased by a factor of three from the time of the fly-by until
$10^8$ years later.

\item If the intruder in a fly-by is a very low-mass star or a brown dwarf, a sufficiently close fly-by may leave the
intruder bound to the host star, potentially leading to drastic changes in the orbital properties of the planets orbiting
the star. This can significantly increase the effect that such fly-bys have on planetary systems compared to
what we would naively expect; for stellar mass intruders the magnitude of the change in the planetary orbits 
due to the fly-by decreases with decreasing intruder-star mass, suggesting
that sub-stellar mass intruders would do very little damage to planetary systems. We have performed a large
number of scattering experiments between a single-planet system and low-mass intruders to measure the rate at
which the intruder may become bound to the host star. We find that, depending on the velocity dispersion
of the cluster, and given the log-normal IMF of \citet{2001ApJ...554.1274C,2002ApJ...567..304C}, 
up to few per cent of solar-system-like planetary systems may be involved in fly-bys that result
in the capture of the low-mass intruder. Of course, if real clusters have an even larger number of
low-mass intruders than suggested by \citet{2001ApJ...554.1274C,2002ApJ...567..304C}, the 
fraction of systems involved in encounters with such objects will increase.

\item  Fly-bys in general excite the eccentricities of planets left bound to 
either the host or the intruder star, both through direct scattering off the 
intruder, and through any subsequent planet-planet scattering. We find 
that the eccentricity distribution of planets which have been scattered 
inwards due to fly-bys in the 4G system is similar to that of the observed extrasolar planets (see Fig. 7 and  
discussion in section 9.2). Planetary systems where the masses are more
similar (e.g. GM \& 4J) tend to overproduce eccentric planets compared
to the observed extrasolar planets.

\item  Planet-planet scattering, triggered either by fly-bys or by primordial 
instability decreases the semi-major axes of one or more planets while leaving 
others on either much wider orbits, or unbinds them entirely. In our 
simulations of the four gas giants, we find that planets are only
rarely scattered onto orbits tighter than 4 au, which is not consistent
with the semi-major axes of observed planets. To reproduce the extrasolar 
planet population it is thus needed to allow for disc migration before the
fly-by, allowing for planets to be on orbits inside the ice-line.

\item  During the phase of planet-planet scattering, one or more 
planets are successively scattered onto wider and wider orbits
until they become unbound. Such planets can,
since they are so far from the host star, be observed using imaging techniques.

\end{enumerate}

While not studied in this paper \citep[but see][]{2007MNRAS.377L...1M, 2009MNRAS.394L..26M} another important type of
encounters is binary-single encounters, in which a single star encounters
a stellar binary system. Such encounters can, depending on the masses
of the stars, leave the initially single star in the binary, replacing one of
the original binary components. The key point here is that in such binaries, any
planetary system will be randomly oriented with respect to the stellar
companion, and as such will often be highly inclined with respect to the
latter. In highly inclined binaries ($i> 39.2^{\circ}$) the Kozai mechanism
operates, exciting the eccentricities of the planets and thereby potentially
triggering planet-planet scattering.

We have combined our results of how fly-bys
affect planetary systems with the rate at which 
encounters in young stellar clusters occur, to 
measure how  important such encounters are. 
As an example, we can look at the effects of
fly-bys on the four gas giants of the solar system.
We find that in between 5 and 15 per cent of such systems,
fly-bys will trigger planet-planet scattering, leading to the
ejection of one or more planets. The fraction
of systems from which planets are ejected varies slowly
with the number of stars in the cluster, $N$. More
stars are believed to form in low-$N$ clusters than
in more rich clusters. Hence, the contribution from 
rather small clusters to the field
population of planetary systems that have been
made unstable by fly-bys, can be significant.
An important point regarding the fraction of systems
that are affected by fly-bys concerns the properties of
the systems before any stellar encounters. To explain
the observed semi-major axes distribution many systems
must initially have contained gas giants on much tighter
orbits than what is seen in the solar system. The effects
of fly-bys on such systems have not yet been explored,
but will be the topic of a forthcoming paper.

In summary, fly-bys in young clusters will significantly change
the properties of the population of planetary systems which
initially resemble the solar system.

\section*{Acknowledgements}
The simulations performed in this paper using NBODY6 and
MERCURY6 were carried out on computer 
hardware which was purchased with grants from the the Royal 
Physiographic Society in Lund. To measure the capture rates
of sub-stellar mass objects by planetary systems, we used the
Starlab package, whose principal contributors are 
Piet Hut, Steve McMillan, Jun Makino and Simon Portegies Zwart.
We thank Ross Church for his careful reading of the manuscript
and for helping to improve it.

\bibliography{Malmberg10a}
\bibliographystyle{mn2e} 

\appendix
\section{Analytical formulae predicting the effects of fly-by's}

In order to write down the equations we use the standard Keplerian orbital
elements describing the shape of the orbits, as well
as the orientation of the intruder star's orbit with respect to the orbit of the planet.

Let, $a_{\rm p}$ be the semi-major axis and $e_{\rm p}$ the
eccentricity of the planet long before the encounter while $e_{\rm I}$ is the 
eccentricity of the intruder star in its hyperbolic orbit around the barycentre
of the planetary system.

In order to define the orientation angles we define the line of nodes as the line
formed by the intersection of the orbital planes of the planet and the
intruder star. The ascending node is the point
on this line at which the intruder star cross the orbital plane of the planet, 
measured in the orbital plane of the planet in the the same direction as
the angular momentum vector of the planetary system. The longitude
of the ascending node, $\Omega$ is measured from the pericentre
of the planet in its orbital plane. Furthermore, $i$
is the inclination between the two orbital planes and $\omega$ is the longitude
of pericentre of the intruder star, measured in its orbital plane from the
ascending node, in the direction of its motion around the planetary system.

Finally, we define $m_{\rm H}$ to be the mass of the host star, $m_{\rm p}$
to be the mass of the planet, $m_{\rm I}$ to be the mass of the intruder 
star and $r_{\rm min}$ to be the minimum separation between the
stars in the fly-by. We write the sum of the mass of the planetary system (planet
and host star) as $M_{\rm ps}$ and the sum of the mass of the 
planetary system and the intruder star as $M_{\rm tot}$.
 
For planets on initially eccentric orbits the change in eccentricity, $\delta e$ is
\citep{1996MNRAS.282.1064H}:

\begin{multline} \label{eq:eccOrbit}
\delta e = -\frac{15}{4} \frac{m_{\rm I}}{\sqrt{M_{\rm ps 
} M_{\rm tot}}} 
 \left( \frac{a_{\rm p}}{r_{\rm min}} \right)^{3/2}
\frac{e_{\rm p} \sqrt{1-e_{\rm p}^2}} {(1+e_{\rm I})^{3/2}} \\ \times 
\left \{
  \sin^2{i} \sin{2 \Omega} 
\left [     \arccos(-1/e_{\rm I}) + \sqrt{e_{\rm I} ^2 - 1}  \right ]
+ \parenthnewln
\frac{1}{3} \left[ (1+\cos^2{i}) \cos{2 \omega} \sin{2 \Omega} + \parenthdoublenewln
2 \cos{i} \sin{2 \omega} \cos{2 \Omega}  \right] 
\frac{(e_{\rm I}^2-1)^{3/2}}{e_{\rm I}^2} \right \}.
\end{multline}

This result vanish if the eccentricity of the planet equals zero, which is why one
must include higher order terms in the derivation. For a distant encounter the change
in eccentricity for a planet on an initially circular orbit is \citep{1996MNRAS.282.1064H}:

\begin{multline} \label{eq:circOrbitDist}
\delta e = \frac{15}{8} \frac{ m_{\rm I} \left| m_{\rm H} - m_{\rm p} \right| }{ M^{2}_{ps} }  
\left(   \frac{M_{\rm ps}}{M_{\rm tot}} \right)^{1/2} \left(  \frac{a_{\rm p}}{r_{\rm min}}\right)^{5/2} \\
\frac{1}{e_{I}^{3} (1+e_{\rm I})^{5/2}} \times \left \{  \cos^2{i} \sin^2{\omega}
\left[  f_1 (e_{\rm I}) (1 - \frac{15}{4}\sin^2{i}) \parenthdoublenewln
+ \frac{2}{15} (e_{\rm I}^2-1)^{5/2} ( 1-5 \sin^2{\omega} \sin^2{i} )  \right]^2  \parenthnewln
+ \cos^2{\omega} \left[  f_1 (e_{\rm I}) (1 - \frac{5}{4}\sin^2{i}) \parenthdoublenewln
+ \frac{2}{15} (e_{\rm I}^2-1)^{5/2} ( 1-5 \sin^2{\omega} \sin^2{i} )  \right]^{2}
\right \}^{1/2},
\end{multline}
where

\begin{equation}
f_1({e_{\rm I}}) = e_{\rm I}^4 \arccos(-1/e_{\rm I}) + \frac{\sqrt{e_{\rm I}^2-1}}{15} (-2+9 e_{\rm I}^2
+ 8 e_{\rm I}^4)
\end{equation}

Finally, for tight encounters involving planets on initially circular orbits, the change in eccentricity
is \citep{1996MNRAS.282.1064H}:

\begin{multline} \label{eq:circOrbTight}
\delta e = 3 \sqrt{2 \pi} \frac{m_{\rm I} M_{\rm ps}^{1/4}}{M_{\rm tot}^{5/4}}
\left( \frac{r_{\rm min}}{a_{\rm p}} \right)^{3/4} \frac{ (e_{\rm I} + 1) ^{3/4}} {e_{\rm I}^2} \\
\times \exp \left[ -\left(  \frac{M_{\rm ps}}{M_{\rm tot}}\right)^{1/2} \left( \frac{r_{\rm min}}{a} \right)^{3/2} 
\frac{ \sqrt{e_{\rm I}^2 - 1} - \arccos{(1/e_{\rm I})} }{ (e_{\rm I}-1)^{3/2} } \right] \\
\times \cos^2{\frac{i}{2}} \left[ \cos^4{\frac{i}{2}}  + \frac{4}{9} \sin^4{\frac{i}{2}}  \parenthnewln
+ \frac{4}{3} \cos^2{\frac{i}{2}} \sin^2{\frac{i}{2}} \cos{(4\omega + 2\Omega)}  \right]^{1/2} .
\end{multline}

In fly-bys involving planetary systems containing multiple planets, not only the change in 
eccentricity, $e$, but also the change in inclination, $i$, is important. Thus, we have, using
the same approach as in \citet{1996MNRAS.282.1064H}, derived an equation for the 
change in inclination of the planet, $\delta i_{\rm p}$, with respect to the plane of
its original orbit:

\begin{multline} \label{eq:incChange}
      \sin{(\delta i_{\rm p})} = \frac{3 \pi m_{\rm I}}{8} 
       \sqrt{ \frac {2}{M_{\rm tot} M_{\rm ps}  (1-e_{\rm p}^2)}}
       \left( \frac{a_{\rm p}}{r_{\rm min}} \right)^{3/2} \\
       \times  \sin{i} \cos{i}  \left [  (1+3e_{\rm p}^2)^2 \sin^2{\Omega}
       + (1-e_{\rm p}^2)^2 \cos^2{\Omega} \right ]^{1/2}
\end{multline}

This expression is for a fly-by with the intruder on a parabolic orbit and
has been averaged over the argument of pericentre ($\omega$).

\label{lastpage}

\end{document}